\documentclass[aps,pra,reprint,superscriptaddress,floatfix]{revtex4-2}
\usepackage{float}

\usepackage{graphicx}
\usepackage{dblfloatfix}
\usepackage{placeins}
\usepackage{natbib}

\begin{document}

\title{From Search to GenAI Queries: Global Trends in Physics Information-Seeking Across Topics and Regions}

\author{Yossi Ben-Zion}\email{benzioy@biu.ac.il}\affiliation{Department of Physics, Bar Ilan University, Ramat Gan 52900, Israel}
\author{Omer Michaeli}\affiliation{Department of Physics, Bar Ilan University, Ramat Gan 52900, Israel}
\author{Noah D. Finkelstein}\affiliation{Department of Physics, University of Colorado Boulder, Boulder, Colorado 80309, USA}

\date{\today}

\begin{abstract}
The emergence of generative artificial intelligence (GenAI) marks a potential inflection point in the way academic information is accessed, raising fundamental questions about the evolving role of search in student learning. This study examines this shift by analyzing longitudinal trends in physics-related search and page-view activity, using declines in traditional search behavior as a quantitative proxy for changes in independent information-seeking practices. We analyze Google Trends data for core concepts in Classical Mechanics and Electromagnetism across three academic years (2022–2025) in more than 20 countries, and complement this analysis with Wikipedia page-view data across seven major languages to establish platform independence. The results reveal a substantial, systematic, and persistent global decline in search and page-view activity across most examined physics topics. The magnitude of this decline is domain-dependent, with Mechanics-related content exhibiting sharper and more consistent reductions than Electromagnetism-related content. Pronounced geographic and linguistic heterogeneity is observed: while English-speaking regions show relative stability or only moderate declines, non-English-speaking regions exhibit substantially larger reductions in traditional, search-based information-seeking activity. Despite the overall decrease in volume, the seasonal structure characteristic of academic activity remains robust. Taken together, these findings indicate a redistribution of physics-related information-seeking behavior in academic contexts where generative tools are increasingly available.
\end{abstract}

\maketitle

\section{Introduction}

"People may lie to their families, to their friends, and even to themselves, but they are brutally honest when facing the empty search box." This observation by Stephens-Davidowitz defines the search engine as a "digital truth serum," capturing human intent without filters \cite{stephens2017everybody}. This unique honesty has transformed aggregated search data into a powerful predictive tool, enabling researchers to forecast influenza outbreaks \cite{ginsberg2009detecting}, anticipate stock market volatility with remarkable accuracy \cite{preis2013quantifying,da2011search}, and forecast near-term economic indicators such as unemployment, consumer confidence, and consumption patterns \cite{choi2012predicting}.

In education, search engines have served as a similar diagnostic window for two decades, as students increasingly equate "Googling" with performing research \cite{rowlands2008google}. However, the rise of GenAI marks a critical inflection point. We are witnessing a shift from a model of \textit{finding information} to one of \textit{generating answers}. This study tracks this shift by analyzing the decline in traditional search queries as a quantitative proxy for the changing landscape of one dimension of study habits.

The validity of Google Trends, a publicly available platform that reports relative search volume across time and regions\cite{googletrends2026}, as a proxy for physics-related learning activity rests on distinguishing student search behavior from general public interest. Gillis and Garrison \cite{gillis2022confounding} demonstrated that technical search terms in physics, mathematics, and health sciences exhibit strong "academic cycling", search volumes peak during examination periods, and collapse during breaks, synchronized with semester schedules. This biphasic pattern distinguished technical terminology, searched primarily by students, from lay terms searched by the general public, which did not show such periodicity.

Arslan et al. \cite{arslan2020using} extended this finding globally, showing that computer science search patterns across more than 120 countries track academic calendars worldwide, with COVID-19 school closures providing causal evidence: when schools closed abruptly in March 2020, technical searches collapsed then recovered with remote instruction. Together, these findings establish that Google Trends data for physics concepts captures classroom-driven learning tools synchronized with formal instruction.

The validity of search data as a proxy for authentic, moment-of-need information-seeking receives striking empirical support from an unexpected source: academic cheating. During the first wave of emergency online education in Spring 2020, Bilen and Matros \cite{bilen2021online} documented a striking natural experiment. As the College Board administered unproctored online Advanced Placement examinations, the researchers tracked Google Trends data at hourly resolution across the United States. The moment the AP Physics C: Mechanics exam commenced, searches for core mechanics concepts surged by more than an order of magnitude. This spike exhibited precise temporal synchronization with the examination window, collapsing back to baseline immediately after submission deadlines. The hourly accuracy of this correspondence demonstrates that students search when stuck.

While the adoption of GenAI is a global trend, with roughly one in six people worldwide now utilizing these tools \cite{microsoft2026global}, studies show that levels of usage \cite{misra2025measuring}, trust \cite{edelman2025trust}, and infrastructure vary significantly between countries \cite{microsoft2026global}. These differences, particularly among Gen Z students \cite{oecd2025experience}, may be key factors in how information-seeking habits change in the classroom. 

Recent studies in Physics Education Research (PER) have examined GenAI across several distinct areas. This research has characterized AI as a tool for automated grading and assessment \cite{PhysRevPhysEducRes.20.020144,PhysRevPhysEducRes.21.010126,PhysRevPhysEducRes.21.010128,PhysRevPhysEducRes.21.010136}, measured its problem-solving performance on conceptual and representational tasks \cite{6fmx-bsnl, PhysRevPhysEducRes.20.010109,PhysRevPhysEducRes21010154, PhysRevPhysEducRes.21.010153}, and explored its potential as a collaborative partner in laboratory or classroom settings \cite{PhysRevPhysEducRes.21.010149, ggy1-3kjk, PhysRevPhysEducRes.21.010147}. Additionally, studies have documented instructor adoption of these tools and their utility in instructional task development \cite{PhysRevPhysEducRes21010154, 10.1119/5.0252343, r2fn-kdy4, PhysRevPhysEducRes.19.020128,kieser2023educational}. While these investigations provide data on AI’s capabilities within formal instruction, a gap remains regarding broader shifts in student behavior. Current research has primarily focused on the tool's pedagogical efficacy, leaving the fundamental changes in independent study practices and the displacement of established resources such as Wikipedia largely unexplored. Furthermore, existing studies have not systematically quantified whether this impact varies between distinct knowledge domains, specifically, between the intuitive nature of Mechanics and the abstract formalism of Electromagnetism. Finally, the magnitude of this behavioral change has not been directly compared across different countries and cultures.

\section{Research Questions}

This study investigates patterns of change in the information-seeking behaviors of physics students following the emergence of GenAI, utilizing large-scale search data as a proxy for learning engagement. Specifically, we address the following three research questions:

\begin{enumerate}

    \item How do longitudinal trends in the use of search engines as a physics learning tool change following the emergence of Generative Artificial Intelligence, particularly in terms of search volume and seasonal periodicity?
    \item To what extent do changes in search patterns differ between the introductory physics domains of Classical Mechanics and Electromagnetism following the emergence of GenAI?
    \item How does the magnitude of observed changes in search behavior vary across different countries and cultures?

\end{enumerate}

\section{Method}

\subsection{Data Source and Instrument Characteristics}
In this study, we utilized the Google Trends platform to quantify longitudinal information-seeking behavior. Unlike server logs that report absolute query counts, Google Trends provides a normalized metric known as Relative Search Volume (RSV). For a specific topic $k$, within a geographic region $r$, and within a specific time window $T$, the reported value $RSV(k, r, t)$ at any given time point $t \in T$ is calculated based on the ratio of specific queries to the total search activity, normalized against the peak activity within that period:

\begin{equation}
    RSV(k, r, t) = 100 \times \frac{ \left( \frac{n(k, r, t)}{N(r, t)} \right) }{ \max_{\tau \in T} \left( \frac{n(k, r, \tau)}{N(r, \tau)} \right) }
\end{equation}

where $n(k, r, t)$ denotes the absolute number of queries for topic $k$, and $N(r, t)$ represents the total search volume (across all categories) in that region. The denominator represents the maximum relative search probability observed throughout the time window $T$, where the variable $\tau$ iterates over all time points in the set $T$. This scaling ensures that the relative interest peak corresponds to a value of 100.

Beyond the basic normalization, each query can be further refined using two primary instrument characteristics: \textbf{Topic entities} and \textbf{Category filters}. These features allow for a more nuanced extraction of search data by shifting the focus from literal string matching to semantic intent and contextual relevance.

\begin{itemize} \item \textbf{Topic Entities:} Unlike literal "Search Terms," a Topic entity represents a semantic concept that aggregates various synonyms, misspellings, and translations across different languages into a single unit of analysis. \item \textbf{Category Filters:} This feature restricts the search volume to a specific domain (e.g., Science, Finance, or Sports), ensuring that the captured interest is contextualized within a relevant field and excluding unrelated colloquial or commercial activity. \end{itemize}

To isolate academic intent and resolve semantic ambiguity, data queries were filtered using the broad "Science" category classification. While a more granular "Physics" sub-category exists, we deliberately opted for the broader classification to maximize signal fidelity and minimize noise artifacts associated with data sparsity in lower-volume regions. A preliminary sensitivity analysis conducted on a representative subset of topics (spanning 8 courses across 3 academic years, yielding 24 paired comparisons) revealed a strong correlation ($r > 0.95$) between data filtered by "Science" and "Physics," confirming that the broader categorization preserves domain specificity while ensuring robust statistical power. This filtering is essential for polysemous topics, ensuring that queries for ambiguous concepts such as \textit{Energy} or \textit{Force} are contextualized within scientific inquiry rather than general contexts.

Furthermore, to ensure the data captures conceptual interest independent of syntax or native language, acquisition relied on Google's "Topic" entities rather than specific "Search Terms." The underlying algorithm maps queries across various alphabets and phrasings into single semantic entities. For instance, the topic entity for \textit{Kinetic Energy} aggregates queries such as \textit{Kinetic energy} (English) and \textit{Kinetische Energie} (German) into a single unit of analysis. This linguistic invariance allows for a valid cross-regional comparison of epistemic engagement with physical concepts.

\subsection{Cross-Regional Normalization of Search Interest}

Google Trends enables cross-regional comparisons of search interest through a normalized measure of Relative Search Volume (RSV). Due to platform constraints, direct cross-regional normalization is limited to a small number of regions (up to five) per query. To compare relative attention density across regions, we define a cross-regional normalization in which all observations are scaled to a common global maximum:

\begin{equation}
RSV_{\text{region}}(k, r_i, t) = 100 \times 
\frac{ \left( \frac{n(k, r_i, t)}{N(r_i, t)} \right) }
{ \max_{r \in R, \tau \in T} \left( \frac{n(k, r, \tau)}{N(r, \tau)} \right) }
\label{eq:cross_regional_rsv}
\end{equation}

Here, $n(k, r_i, t)$ denotes the number of queries for topic $k$ in region $r_i$ at time $t$, and $N(r_i, t)$ represents the total search volume in that region. This formulation normalizes search interest by overall search activity and expresses it relative to the highest observed attention density across all regions and times. As a result, the metric captures relative attention within each region rather than absolute query counts, mitigating biases associated with population size and internet penetration.

\subsection{Temporal Scope}

The temporal window of this study encompasses three complete academic cycles, spanning from August 15, 2022, to August 14, 2025. This delineation corresponds to the standard academic calendar of the Northern Hemisphere, which encompasses the primary selected regions (United States and India) and dominates the global aggregate of academic activity. 

Methodologically, the analysis does not attempt to align academic calendars across countries. Instead, it examines how search behavior changes over time within each country. As a result, differences in academic schedules do not affect the validity of the year-to-year comparisons.

Accordingly, the longitudinal analysis is segmented into three distinct epochs, denoted by their concluding years for analytical brevity:
\begin{enumerate}
\item \textit{Reference Year (‘23’):} (Aug 2022–Aug 2023).
This period serves as a baseline, capturing search behavior prior to the widespread academic adoption of generative AI tools.

\item \textit{Transition Year (‘24’):} (Aug 2023–Aug 2024).
This phase corresponds to the rapid diffusion and experimental incorporation of large language models into student learning practices.

\item \textit{Integration Year (‘25’):} (Aug 2024–Aug 2025).
This epoch reflects a period in which GenAI tools have become more established and routinely integrated into the academic landscape.

\end{enumerate}

\subsection{Topic Selection and Taxonomy}
To ensure the pedagogical validity of the search queries, topics were selected through a systematic curation process based on the table of contents of a seminal introductory physics textbook \cite{shankar2016fundamentals,shankar2019fundamentals}. The study scope was strictly limited to two foundational domains: \textit{Classical Mechanics} and \textit{Electromagnetism}.

The initial pool of topics underwent a rigorous filtration process based on two exclusion criteria. First, we enforced Domain Specificity by excluding auxiliary mathematical concepts often appearing in physics curricula, such as \textit{Vector} or \textit{Derivative}, to ensure that the measured search volume reflects engagement with physical principles rather than general mathematical tools. Second, we assessed statistical viability through a data quality control process in which all candidate time series were inspected for suitability for longitudinal analysis. Topics exhibiting extended intervals of zero values or repeated sharp collapses to zero were excluded, as such patterns undermine the stability and interpretability of growth rates and seasonal correlations across academic cycles.

\subsection{Geographic Sampling}

We defined our initial sampling frame based on the 40 most populous countries worldwide \cite{un_wpp_2024}, corresponding to a population threshold of approximately 38 million people. This cutoff was chosen to ensure coverage of the majority of the global student population while reducing the likelihood of noisy or sparse search data, which are more common in smaller countries.

From this initial pool, countries were included in the final sample based on the availability, consistency, and signal quality of the search data. In several highly populated countries, search activity for specific physics-related topics was insufficient, with frequent weeks exhibiting zero search volume, leading to an inadequate signal-to-noise ratio. Such cases were excluded from the analysis. In addition, regions such as mainland China were excluded because Google is not the dominant search platform, in order to mitigate systematic platform bias.

Beyond population size, we explicitly considered geographic diversity and data richness. Saudi Arabia and Malaysia were included despite falling slightly below the population threshold, as their inclusion improves regional representation and because search activity from these countries exhibited consistently high volume and low noise across the analyzed queries.

Australia was retained as an intentional exception to strengthen the English-speaking reference group for cross-linguistic comparison. The resulting dataset consists of more than 20 countries and provides a diverse cross-sectional view of how students in different regions are shifting from traditional search engines toward GenAI tools.

\subsection{Statistical Protocol}

To ensure technical consistency across the three-year period, data for each query were retrieved via a single continuous search spanning the entire duration. Following the logic in Eq.~(1), this procedure normalizes all weekly observations relative to a common global maximum ($\tau$) for the entire duration, preventing mathematical artifacts associated with independent annual rescaling. This process yielded $N=53$ weekly Relative Search Volume (RSV) data points for each academic cycle: Reference ('23), Transition ('24), and Integration ('25).

For each search query, we define the annual mean RSV ($\mu_y$) as:
\begin{equation}
\mu_y = \frac{1}{N} \sum_{i=1}^{N} RSV_{i,y}
\end{equation}
where $i$ represents the week and $y$ denotes the academic year. To quantify the shift between the baseline and the stabilized AI-integration period, we calculated the dimensionless percentage change ($\%\Delta$):
\begin{equation}
\%\Delta = \left( \frac{\mu_{25} - \mu_{23}}{\mu_{23}} \right) \times 100
\end{equation}
By using annual means rather than weekly raw values, we stabilize the metric against short-term volatility. As a dimensionless ratio, $\%\Delta$ allows for a direct comparison between topics with different baseline search volumes, providing a normalized measure of the relative shift in search interest.

To determine whether the observed changes represent systematic shifts rather than stochastic noise, we employed paired-sample $t$-tests. The choice of a paired analysis is contextually motivated by the temporal structure of the academic calendar; by matching corresponding weeks across cycles, we effectively control for the periodic fluctuations inherent to academic activity, such as examination surges and holiday breaks. While search data often exhibit non-normal distributions due to these periodic spikes, with $N=53$ paired observations the paired-sample $t$-test is generally robust to moderate deviations from normality in the distribution of the paired differences, making it an appropriate tool for assessing the significance of the mean difference.

For clarity of interpretation, positive values of the $t$ statistic and positive values of Cohen's $d$ indicate an increase in search interest from the 2023 reference year to the 2025 integration year, whereas negative values indicate a decline. This convention aligns with the percentage change $\%\Delta$, defined as $(\mu_{25}-\mu_{23})/\mu_{23}$, ensuring consistent interpretation across all reported metrics.

Given that $p$-values can be sensitive to sample characteristics, we prioritize Cohen’s $d$ for paired samples to interpret the practical significance of the observed shifts. We define $d$ as the mean difference divided by the standard deviation of the differences. Following standard conventions in educational research, we interpret values of $0.2$, $0.5$, and $0.8$ as small, medium, and large effects, respectively. This metric allows us to distinguish between results that are merely statistically significant and those that represent a substantial change in student information-seeking behavior.

Weekly search time series exhibit seasonal structure and temporal dependence between adjacent observations, which may relax the strict independence assumption of week-by-week comparison tests. However, the goal of the present analysis is not to model short-term temporal dynamics but to identify systematic shifts in the mean level of search activity across complete academic cycles. Accordingly, while weekly autocorrelation may affect the precise estimation of pointwise statistical significance, it is not expected to materially affect the direction or the order of magnitude of the reported effects, which are primarily interpreted through standardized effect sizes.

Finally, for comparisons across different countries and physics domains, we examine the distribution of outcomes rather than aggregate global means. Because RSV is normalized within each query but not across different queries, it is mathematically invalid to average raw RSV values from disparate topics (e.g., comparing \textit{Force} directly to \textit{Gauss's Law}). Instead, we evaluate the median $\%\Delta$ and the distribution of Cohen’s $d$ values within each category. This approach provides a more accurate cross-sectional view, identifying whether the observed trends are consistent across an entire conceptual domain or driven by specific outlier topics.

All data processing and statistical analyses were conducted in Python. We utilized the \texttt{NumPy} and \texttt{Pandas} libraries for data management, while \texttt{SciPy} was employed for all statistical procedures, including paired-sample $t$-tests and correlation analyses (Pearson and Spearman rank correlations). Statistical significance was defined at the $\alpha = 0.05$ level. Throughout the results section, significance levels are indicated using the following notation: $^{*}p < 0.05$, $^{**}p < 0.01$, and $^{***}p < 0.001$.

\section{Results}
The results section is structured to systematically demonstrate shifts in information-seeking behavior, progressing from foundational validation to global trends. We begin by establishing the link between information-seeking activity and academic instruction by comparing the time series of the topics \textit{Kinetic energy} and \textit{ChatGPT} in the United States, illustrating their synchronization with the academic calendar. Second, we present a macro-level comparison of search volume declines across two groups of physics-related content examined across three primary regions: the United States, India, and the global aggregate. Third, We conduct a granular analysis of four foundational topics (Kinetic energy, Newton's laws of motion, Electric charge, and Electric field) across more than 20 countries. Finally, we provide \textit{Wikipedia} page-view data to verify that the observed search-related trends are independent of the \textit{Google Search} platform.
Throughout this section, references to \textit{Mechanics} and \textit{Electromagnetism} denote groups of physics-related search content.
\subsection{Temporal Validation and Academic Alignment}

\begin{figure}[htbp]
\centering
\includegraphics[width=\linewidth]{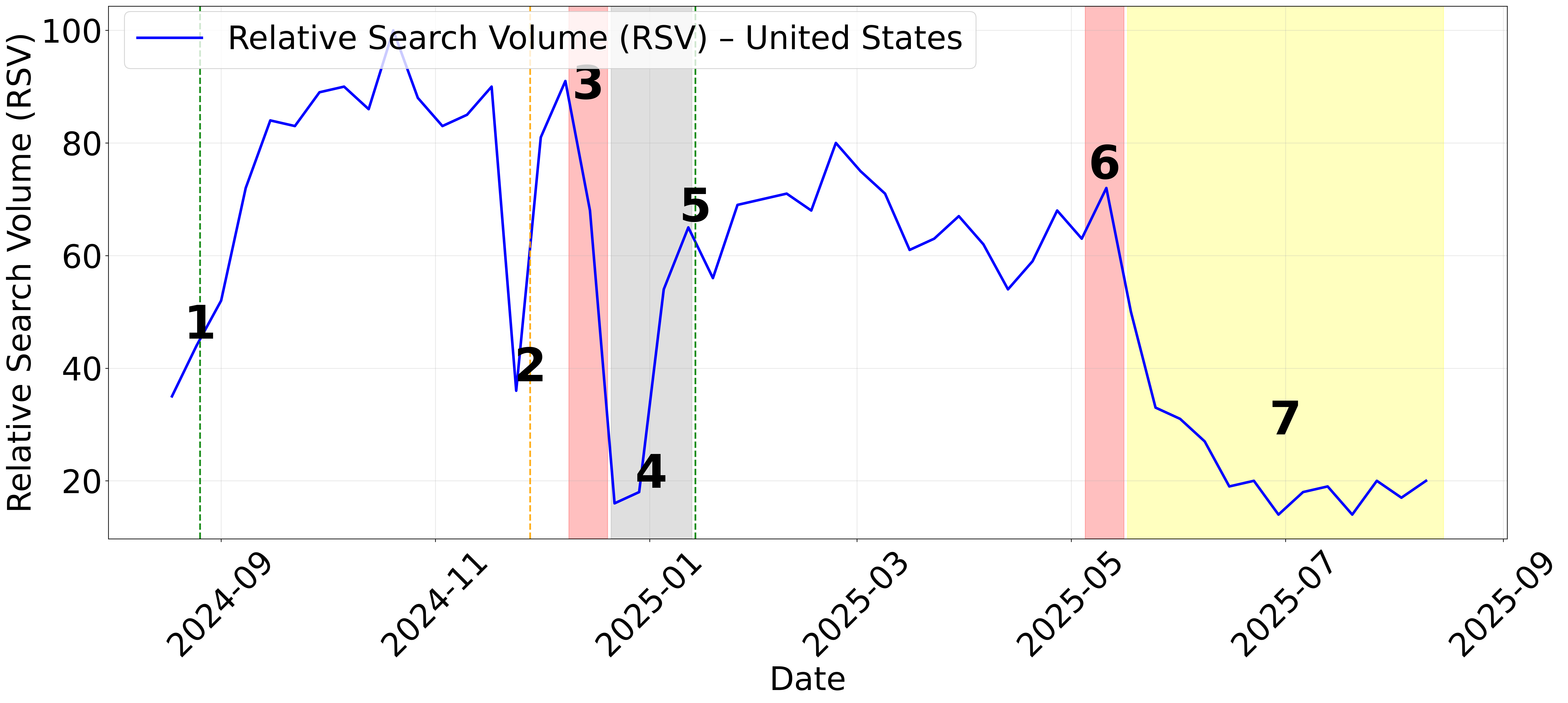}
\caption{\label{fig:kinetic_energy_us}
Temporal distribution of search interest for the topic \textit{Kinetic energy} in the United States over the 2024–2025 academic year, filtered within the \textit{Science} category, with annotated markers corresponding to key academic calendar events. Numbered markers indicate: (1) \textit{Fall semester start}, (2) \textit{Thanksgiving break}, (3) \textit{Fall final examination period}, (4) \textit{Winter break}, (5) \textit{Spring semester start}, (6) \textit{Spring final examination period}, and (7) \textit{Summer break}.
}
\end{figure}

\begin{figure}[htbp]
\centering
\includegraphics[width=\linewidth]{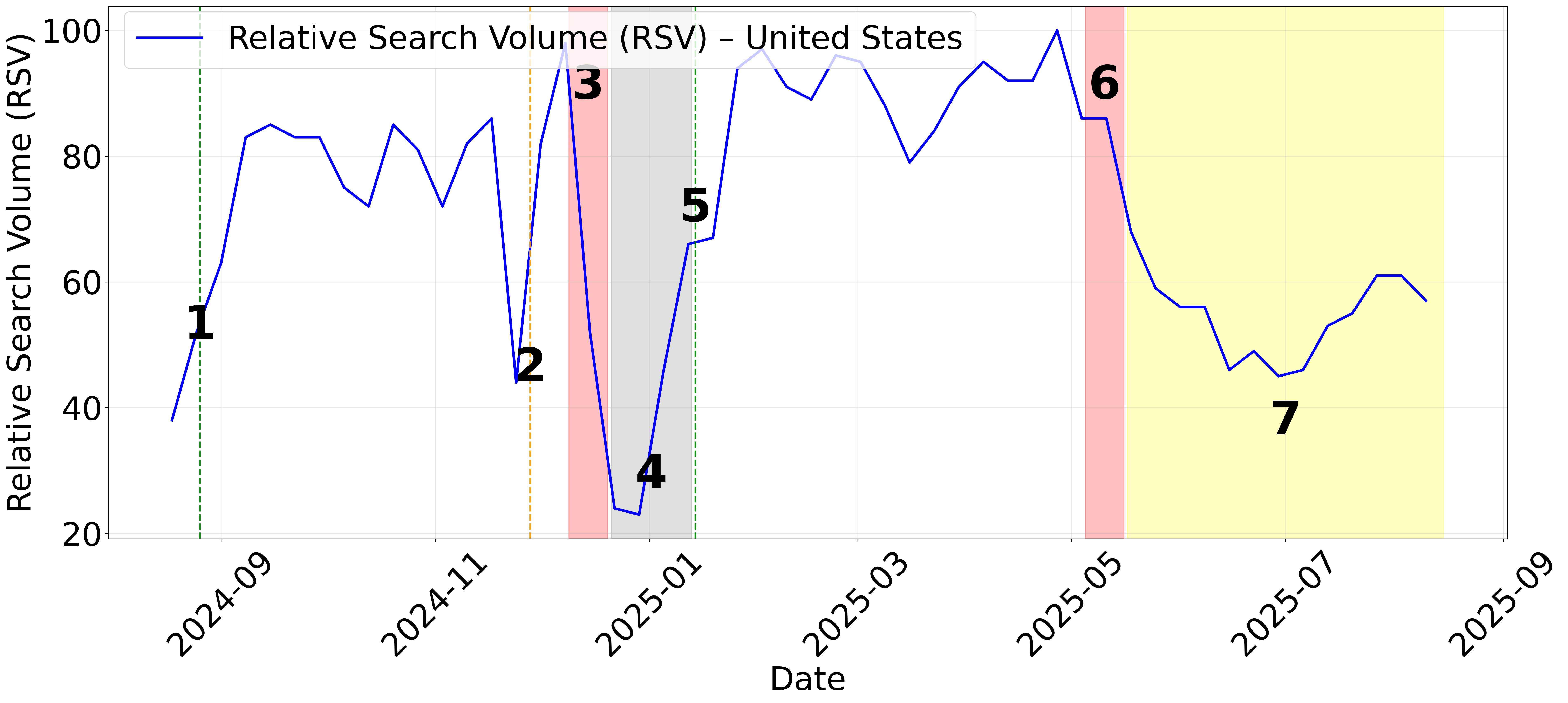}
\caption{\label{fig:chatgpt_us}
Temporal distribution of search interest for the topic \textit{ChatGPT} in the United States over the 2024–2025 academic year, filtered within the \textit{Science} category, with annotated markers corresponding to key academic calendar events that provide contextual reference for potential seasonal patterns. Numbered markers indicate:
(1) \textit{Fall semester start}, 
(2) \textit{Thanksgiving break}, 
(3) \textit{Fall final examination period}, 
(4) \textit{Winter break}, 
(5) \textit{Spring semester start}, 
(6) \textit{Spring final examination period}, and 
(7) \textit{Summer break}.
}
\end{figure}

Figures~\ref{fig:kinetic_energy_us} and~\ref{fig:chatgpt_us} illustrate the relationship between the academic calendar and search activity associated with specific content. As shown in Figure~\ref{fig:kinetic_energy_us}, search interest for the topic \textit{Kinetic energy} in the United States is closely aligned with the academic calendar.

In the same manner, the search interest for the topic \textit{ChatGPT}, filtered within the \textit{Science} category (Fig.~\ref{fig:chatgpt_us}), exhibits clear academic cycling. 

\subsection{Macro-Level Comparison: Content and Regions}

To assess large-scale changes in physics-related search activity, we compared the weekly \textit{Relative Search Volume} (RSV) for content associated with physics instruction between the academic years \textit{2022--2023} and \textit{2024--2025}. We analyze introductory physics-related content across three primary regions: the United States, India, and the global aggregate (\textit{Worldwide}). For content associated with mechanics-related material, data were available for 30 topics worldwide, 27 in the United States, and 26 in India. For content associated with electromagnetism-related material, data were available for 32 topics worldwide, 23 in India, and 17 in the United States. The complete statistical results for all analyzed topics are provided in Tables~\ref{tab:worldwide_mechanics_changes}--\ref{tab:india_em_changes} of the Appendix.

\begin{figure}[htbp]
\centering
\includegraphics[width=\linewidth]{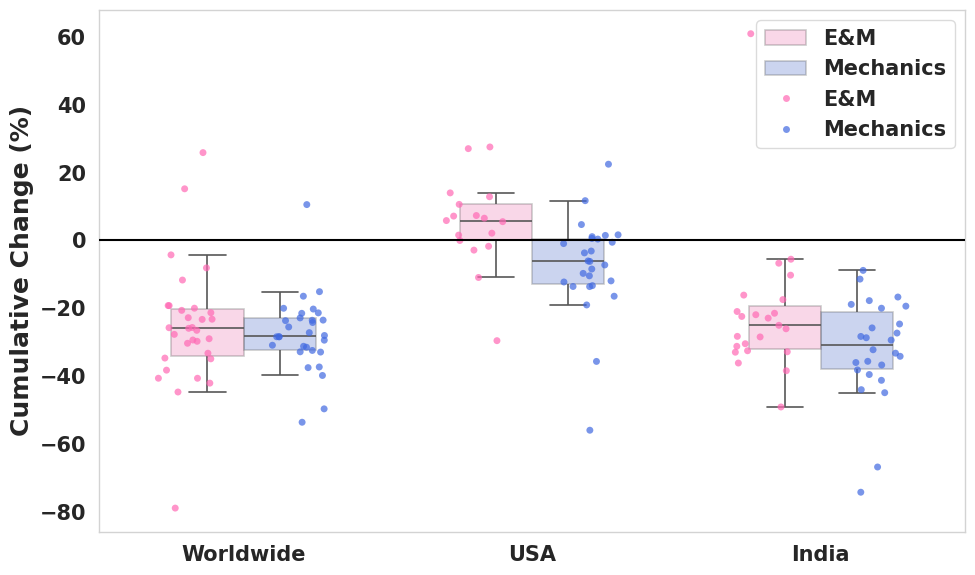}
\caption{\label{fig:box_plot_percent}
Distribution of cumulative percentage changes ($\Delta\%$) in search volume from 2023--2025 across regions and groups of physics-related search content. Box plots represent the median and interquartile ranges, while individual topics are shown as distinct points. A divergence is observed between \textit{Mechanics} (blue) and \textit{Electromagnetism} (pink), particularly in the United States, where \textit{Electromagnetism}-related search content shows a positive median shift. For visualization purposes only, one extreme outlier (+220\% in the worldwide electromagnetism dataset) was excluded to ensure an interpretable scale. Its exclusion does not affect the reported medians or the qualitative conclusions.
}
\end{figure}

\begin{figure}[htbp]
\centering
\includegraphics[width=0.95\linewidth]{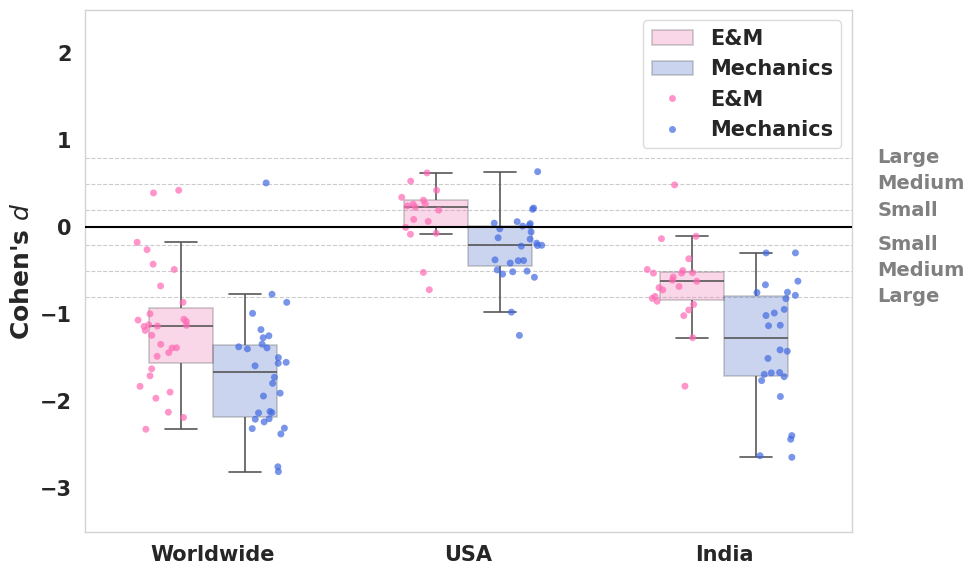}
\caption{\label{fig:box_plot_cohen}
Standardized effect sizes (Cohen's $d$) for physics-related search content across regions. Dashed horizontal lines indicate thresholds for small ($0.2$), medium ($0.5$), and large ($0.8$) effects. In the \textit{Worldwide} and India datasets, the median effect size for \textit{Mechanics}-related search content exceeds the threshold for a large effect ($d > 0.8$), whereas the United States displays substantially smaller effect sizes.}
\end{figure}

Across all tables, paired comparisons reveal large standardized declines for most topics, while Pearson and Spearman correlations remain high, indicating that the academic seasonal structure is preserved despite reductions in overall volume. The distribution of these changes, summarized in Figures~\ref{fig:box_plot_percent} and~\ref{fig:box_plot_cohen}, reveals a substantial global decline in search volume, particularly pronounced in the datasets for the global aggregate and India. The vast majority of analyzed topics exhibit sharp declines with high $t$-statistics and large effect sizes, frequently exceeding $d=1.5$.

In contrast, the United States exhibits a distinct and more heterogeneous pattern. While several mechanics-related search topics show statistically significant declines, their effect sizes are generally smaller and less systematic than those observed worldwide and in India, with some of this content remaining stable. In electromagnetism-related search content, the distribution is shifted toward near-zero or positive cumulative changes, with many topics exhibiting non-significant effects and a positive median.

\begin{figure*}[htbp]
\centering
\makebox[\textwidth][c]{%
  \includegraphics[width=0.95\textwidth]{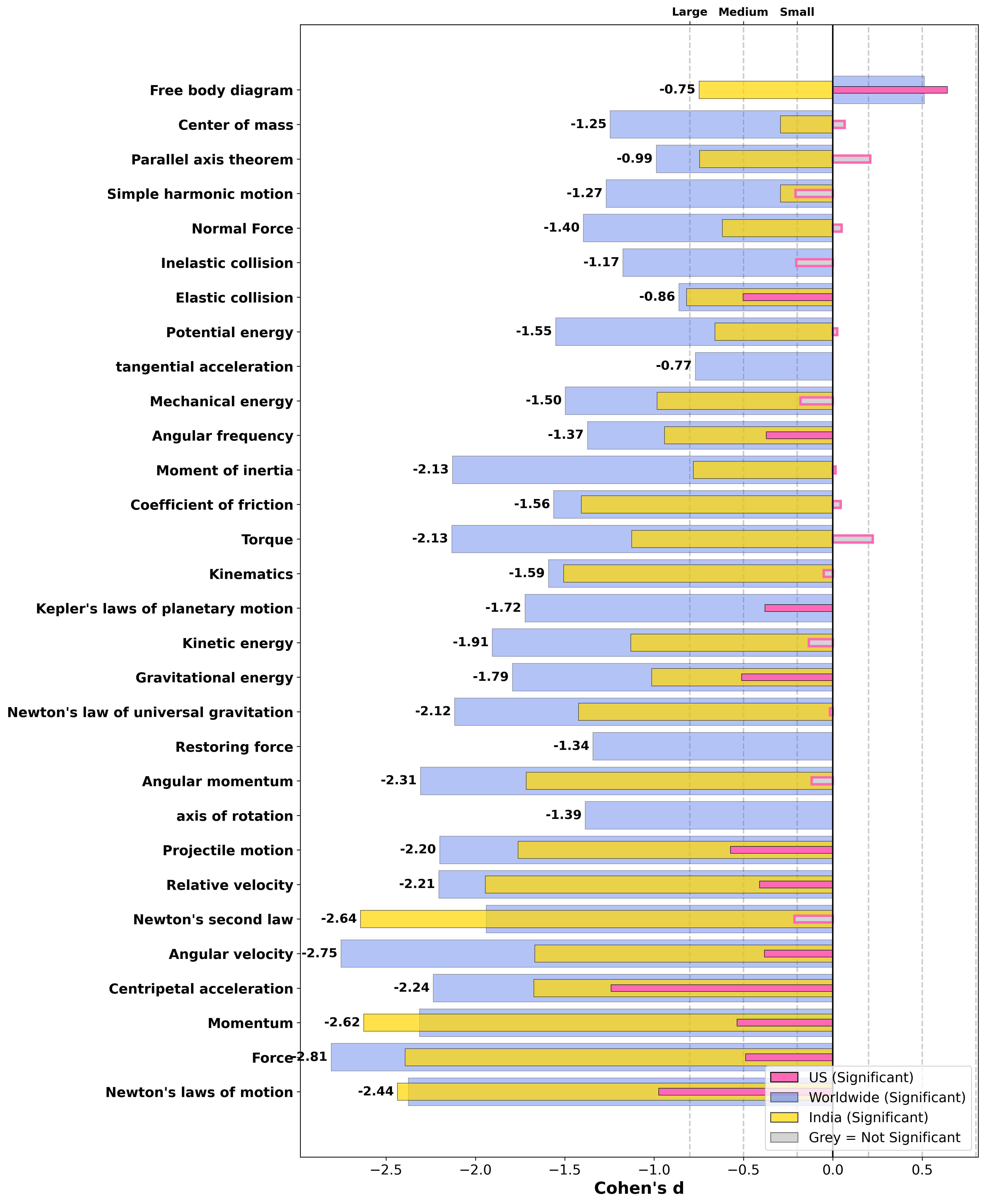}
}
\caption{\label{fig:D_COHEN_Mechanics}
Effect-size (Cohen's $d$) summary for \textit{Mechanics}-related search content, comparing the \textit{2022--2023} reference period with the \textit{2024--2025} integration period across three regions: the United States, \textit{Worldwide}, and India. Bars represent the corresponding Cohen's $d$ values by region. Dashed vertical lines mark the conventional thresholds for small ($d=0.2$), medium ($d=0.5$), and large ($d=0.8$) effects. Statistically significant differences ($p<0.05$) are highlighted in color (pink for the United States, blue for \textit{Worldwide}, and yellow for India), whereas non-significant results ($p\ge 0.05$) are shown in gray.}
\end{figure*}

\begin{figure*}[htbp]
\centering
\makebox[\textwidth][c]{%
  \includegraphics[width=0.95\textwidth]{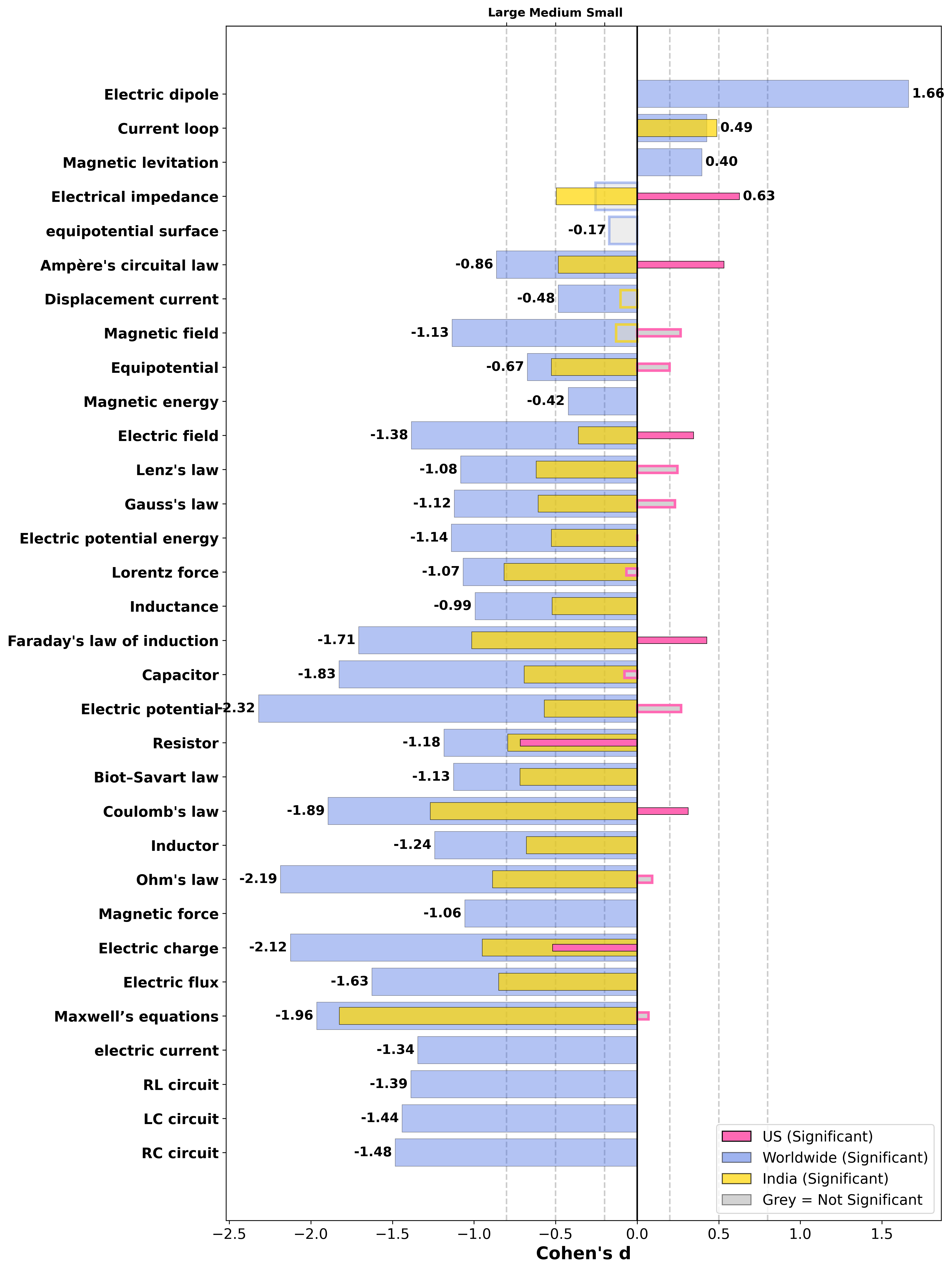}
}
\caption{\label{fig:cohen_d_combined_Electricity}
Effect-size (Cohen's $d$) summary for \textit{Electromagnetism}-related search content, comparing the \textit{2022--2023} reference period with the \textit{2024--2025} integration period across three regions: the United States, \textit{Worldwide}, and India. Bars represent the corresponding Cohen's $d$ values by region. Dashed vertical lines mark the conventional thresholds for small ($d=0.2$), medium ($d=0.5$), and large ($d=0.8$) effects. Statistically significant differences ($p<0.05$) are highlighted in color (pink for the United States, blue for \textit{Worldwide}, and yellow for India), whereas non-significant results ($p\ge 0.05$) are shown in gray.}
\end{figure*}

Alongside these geographical differences, a consistent divergence is observed (Figs.~\ref{fig:box_plot_percent} and ~\ref{fig:box_plot_cohen}), with \textit{Mechanics} showing sharper declines in search volume than \textit{Electromagnetism}. While effect sizes associated with \textit{Mechanics} are substantially larger and predominantly negative, those associated with \textit{Electromagnetism} are notably smaller. Despite the overall decline in search volume, the “academic pulse” remains intact; correlation coefficients ($r$ and $\rho$) remain very high (frequently above $0.8$), indicating that search patterns continue to follow the academic semester. A more granular comparison is provided in Figs.~\ref{fig:D_COHEN_Mechanics} and~\ref{fig:cohen_d_combined_Electricity}, illustrating that the United States exhibits a distinctly different search profile compared to India and the global aggregate, while further highlighting the larger effect sizes observed for \textit{Mechanics} relative to \textit{Electromagnetism}.

\subsection{Global Concept-Specific Analysis}

To examine changes in information-seeking behavior across various countries, we selected four foundational physics-related search topics that are central to introductory physics instruction: two associated with \textit{Mechanics} (\textit{Kinetic energy} and \textit{Newton’s laws of motion}) and two associated with \textit{Electromagnetism} (\textit{Electric charge} and \textit{Electric field}).

These topics were analyzed across a sample of major countries representing broad geographical and linguistic diversity. Data were available for 30 countries for \textit{Kinetic energy}, 28 for \textit{Newton’s laws of motion}, 30 for \textit{Electric charge}, and 24 for \textit{Electric field}, with availability varying by topic due to differences in search volume across regions. The complete statistical results for these topics are provided in Tables~\ref{tab:kinetic_energy}--\ref{tab:electric_field}.

The results for the topic \textit{Kinetic energy}, summarized in Table~\ref{tab:kinetic_energy}, indicate a significant global decline of 28.42\% in search volume, accompanied by a large effect size ($d=-1.91$) and a high Pearson correlation ($r=0.92$). Substantial regional heterogeneity is observed: while the United States shows negligible change ($-3.12\%$, $d=-0.13$) and the United Kingdom exhibits a relatively moderate decline ($-7.89\%$), several other regions show markedly larger reductions in search volume, including South Korea ($-52.02\%$), Ukraine ($-50.19\%$), and Brazil ($-46.46\%$).

\begin{figure*}[htbp]
\centering
\includegraphics[width=\textwidth]{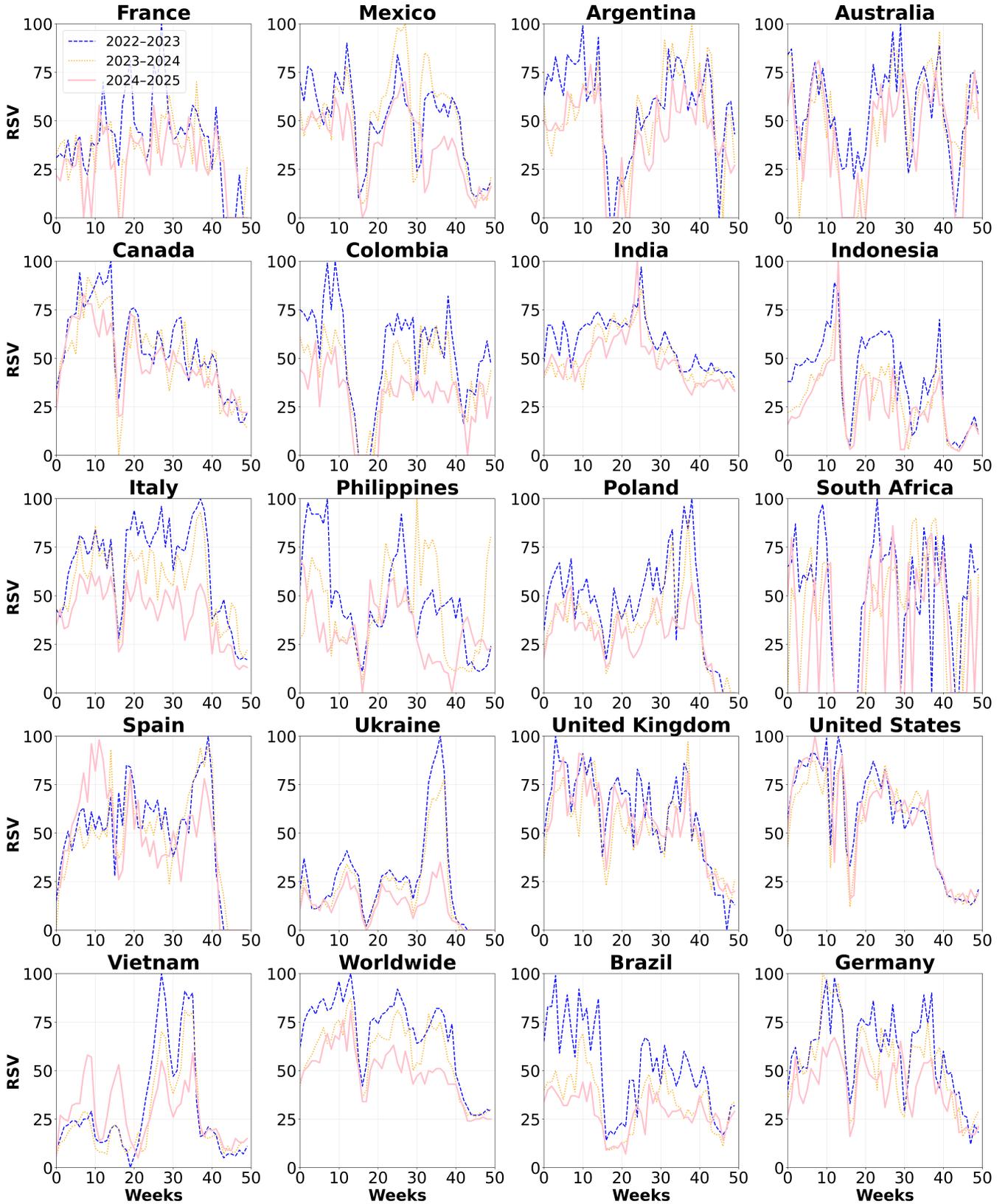}
\caption{\label{fig:kinetic_energy_wide}
Temporal evolution of search interest in \textit{Kinetic energy} across multiple countries, shown as a function of weeks from the start of the academic year, enabling cross-national comparison over three academic years aligned to the academic calendar.}
\end{figure*}

\begin{table}[htbp]
\centering
\caption{\label{tab:kinetic_energy}
Descriptive statistics and effect sizes for \textit{Kinetic energy} search interest across regions (2023--2025)}

\noindent\rule{\linewidth}{0.6pt}

\begin{tabular*}{\linewidth}{@{\extracolsep{\fill}}lcccc}
 & \%$\Delta$ & $t$ & Cohen's $d$ & Pearson $r$ \\\hline
Worldwide & -28.42 & -13.87*** & -1.91 & 0.92*** \\
Argentina & -24.11 & -6.33*** & -0.87 & 0.75*** \\
Australia & -19.94 & -4.51*** & -0.62 & 0.72*** \\
Brazil & -46.46 & -9.48*** & -1.30 & 0.70*** \\
Canada & -14.52 & -5.52*** & -0.76 & 0.88*** \\
Colombia & -44.16 & -11.31*** & -1.55 & 0.81*** \\
Egypt & -24.95 & -2.75** & -0.38 & 0.55*** \\
France & -32.63 & -5.15*** & -0.71 & 0.61*** \\
Germany & -27.56 & -8.37*** & -1.15 & 0.78*** \\
India & -17.75 & -8.23*** & -1.13 & 0.72*** \\
Indonesia & -36.55 & -7.07*** & -0.97 & 0.75*** \\
Iran & 19.59 & 2.85** & 0.39 & 0.85*** \\
Italy & -36.43 & -11.49*** & -1.58 & 0.81*** \\
Japan & -0.14 & -0.05 & -0.01 & 0.85*** \\
Malaysia & -13.93 & -1.31 & -0.18 & 0.33* \\
Mexico & -24.62 & -6.44*** & -0.88 & 0.76*** \\
Nigeria & 14.86 & 1.70 & 0.23 & 0.55*** \\
Pakistan & -29.59 & -5.82*** & -0.80 & 0.10 \\
Philippines & -25.87 & -3.22** & -0.44 & 0.25 \\
Poland & -38.39 & -8.41*** & -1.16 & 0.85*** \\
Russia & -21.04 & -7.17*** & -0.98 & 0.91*** \\
Saudi Arabia & -17.63 & -1.22 & -0.17 & 0.50*** \\
South Africa & -32.10 & -3.80*** & -0.52 & 0.48*** \\
South Korea & -52.02 & -5.75*** & -0.79 & 0.39** \\
Spain & -8.52 & -1.53 & -0.21 & 0.75*** \\
Thailand & -36.74 & -6.72*** & -0.92 & 0.88*** \\
Turkey & -26.62 & -2.17* & -0.30 & 0.44*** \\
Ukraine & -50.19 & -5.20*** & -0.71 & 0.84*** \\
United Kingdom & -7.89 & -2.47* & -0.34 & 0.82*** \\
United States & -3.12 & -0.98 & -0.13 & 0.87*** \\
Vietnam & -6.22 & -0.63 & -0.09 & 0.64*** \\
\end{tabular*}

\noindent\rule{\linewidth}{0.6pt}

\end{table}
Figure~\ref{fig:kinetic_energy_wide} provides a longitudinal perspective on these search-related shifts across twenty countries. The visualization shows that the declines observed in most nations are consistent throughout the examined period and are not readily attributable to short-term stochastic variation. Furthermore, the figure illustrates that successive academic cycles retain their seasonal structure, while absolute search volumes progressively diminish year over year. The distinct patterns observed in outliers, such as the United States and the United Kingdom, which maintain relatively stable search levels, are also clearly visible in this cross-national comparison.

\begin{table}[htbp]
\centering
\caption{\label{tab:newtons_laws}
Descriptive statistics and effect sizes for \textit{Newton's laws of motion} search interest across regions (2023--2025)}

\noindent\rule{\linewidth}{0.6pt}

\begin{tabular*}{\linewidth}{@{\extracolsep{\fill}}lcccc}
 & \%$\Delta$ & $t$ & Cohen's $d$ & Pearson $r$ \\\hline
Worldwide & -39.83 & -17.29*** & -2.37 & 0.84*** \\
Argentina & -34.32 & -8.70*** & -1.19 & 0.72*** \\
Australia & -34.23 & -8.55*** & -1.18 & 0.78*** \\
Brazil & -39.83 & -10.50*** & -1.44 & 0.69*** \\
Canada & -36.46 & -8.72*** & -1.20 & 0.76*** \\
Colombia & -33.20 & -7.47*** & -1.03 & 0.68*** \\
Egypt & -41.29 & -5.31*** & -0.73 & 0.39** \\
France & -7.59 & -1.25 & -0.17 & 0.62*** \\
Germany & -28.38 & -7.03*** & -0.97 & 0.66*** \\
India & -36.01 & -17.73*** & -2.44 & 0.89*** \\
Indonesia & -52.92 & -8.97*** & -1.23 & 0.57*** \\
Iran & 4.37 & 0.32 & 0.04 & 0.64*** \\
Italy & -49.50 & -9.00*** & -1.24 & 0.56*** \\
Japan & -23.17 & -3.16** & -0.43 & 0.49*** \\
Malaysia & -50.12 & -6.76*** & -0.93 & -0.05 \\
Mexico & -43.42 & -7.66*** & -1.05 & 0.69*** \\
Nigeria & -39.71 & -9.54*** & -1.31 & 0.44*** \\
Pakistan & -55.15 & -15.61*** & -2.14 & 0.20 \\
Philippines & -55.50 & -6.92*** & -0.95 & 0.49*** \\
Poland & -32.59 & -5.84*** & -0.80 & 0.74*** \\
Russia & -9.27 & -2.86** & -0.39 & 0.96*** \\
Saudi Arabia & -27.94 & -2.48* & -0.34 & 0.51*** \\
South Africa & -40.25 & -7.50*** & -1.03 & 0.87*** \\
Spain & -5.39 & -0.86 & -0.12 & 0.58*** \\
Thailand & -8.34 & -2.00 & -0.27 & 0.86*** \\
Ukraine & -25.03 & -6.64*** & -0.91 & 0.95*** \\
United Kingdom & -33.32 & -6.54*** & -0.90 & 0.56*** \\
United States & -35.68 & -7.09*** & -0.97 & 0.80*** \\
Vietnam & -75.32 & -7.62*** & -1.05 & 0.80*** \\
\end{tabular*}

\noindent\rule{\linewidth}{0.6pt}

\end{table}
Table~\ref{tab:newtons_laws} presents the results for the topic \textit{Newton’s laws of motion}. The global decline in search volume is 39.83\%, accompanied by a large effect size ($d=-2.37$), which is larger than that observed for the topic \textit{Kinetic energy}. In the United States, search volume declined by 35.68\%, contrasting with the relatively stable search pattern observed for this topic in the earlier period. Regional variation in the magnitude of the decline remains evident across the dataset.

\begin{table}[htbp]
\centering
\caption{\label{tab:electric_charge}
Descriptive statistics and effect sizes for \textit{Electric charge} search interest across regions (2023--2025)}

\noindent\rule{\linewidth}{0.6pt}

\begin{tabular*}{\linewidth}{@{\extracolsep{\fill}}lcccc}
 & \%$\Delta$ & $t$ & Cohen's $d$ & Pearson $r$ \\\hline
Worldwide &-25.67 & -15.47*** & -2.12 & 0.88*** \\
Argentina & -56.13 & -14.17*** & -1.95 & 0.23 \\
Australia & -8.31 & -2.84** & -0.39 & 0.88*** \\
Brazil & -34.77 & -10.04*** & -1.38 & 0.87*** \\
Canada & -21.63 & -11.04*** & -1.52 & 0.96*** \\
Colombia & -52.98 & -15.91*** & -2.19 & 0.81*** \\
Egypt & -4.18 & -0.53 & -0.07 & 0.55*** \\
France & -35.05 & -7.36*** & -1.01 & 0.90*** \\
Germany & -22.83 & -8.88*** & -1.22 & 0.85*** \\
India & -20.96 & -6.91*** & -0.95 & 0.46*** \\
Indonesia & -52.83 & -6.27*** & -0.86 & 0.79*** \\
Iran & 19.56 & 2.83** & 0.39 & 0.61*** \\
Italy & -31.03 & -10.64*** & -1.46 & 0.85*** \\
Japan & 4.74 & 1.80 & 0.25 & 0.84*** \\
Malaysia & -28.70 & -6.38*** & -0.88 & -0.26 \\
Mexico & -54.01 & -14.06*** & -1.93 & 0.69*** \\
Nigeria & -50.14 & -11.60*** & -1.59 & 0.40** \\
Pakistan & -43.16 & -13.63*** & -1.87 & 0.45*** \\
Philippines & -45.98 & -5.57*** & -0.77 & 0.37** \\
Poland & -34.68 & -9.16*** & -1.26 & 0.93*** \\
Russia & -21.46 & -5.06*** & -0.69 & 0.87*** \\
Saudi Arabia & -19.90 & -1.62 & -0.22 & 0.34* \\
South Africa & -40.89 & -10.58*** & -1.45 & 0.85*** \\
South Korea & -25.96 & -6.73*** & -0.92 & 0.82*** \\
Spain & -32.28 & -11.60*** & -1.59 & 0.80*** \\
Thailand & -33.38 & -10.48*** & -1.44 & 0.84*** \\
Turkey & -24.65 & -2.45* & -0.34 & 0.26 \\
Ukraine & -36.51 & -8.93*** & -1.23 & 0.93*** \\
United Kingdom & -13.69 & -5.38*** & -0.74 & 0.89*** \\
United States & -10.92 & -3.77*** & -0.52 & 0.92*** \\
Vietnam & 9.67 & 0.86 & 0.12 & 0.38** \\
\end{tabular*}

\noindent\rule{\linewidth}{0.6pt}

\end{table}
Tables~\ref{tab:electric_charge} and~\ref{tab:electric_field} present the results for the topics \textit{Electric charge} and \textit{Electric field}. For the topic \textit{Electric charge}, the global decline in search volume is 25.67\% ($d=-2.12$), and for the topic \textit{Electric field}, it is 25.75\% ($d=-1.38$). In both cases, the United States, the United Kingdom, and Australia exhibit search patterns that differ from the general trend. While most countries show sharp declines in search volume, often exceeding 50\%, these three nations display relatively minor decreases for \textit{Electric charge} (ranging from 8\% to 14\%) and near-stability or slight increases for \textit{Electric field} (+6.57\% in the United States and +5.17\% in the United Kingdom).

\begin{table}[htbp]
\centering
\caption{\label{tab:electric_field}
Descriptive statistics and effect sizes for \textit{Electric field} search interest across regions (2023--2025)}

\noindent\rule{\linewidth}{0.6pt}

\begin{tabular*}{\linewidth}{@{\extracolsep{\fill}}lcccc}
 & \%$\Delta$ & $t$ & Cohen's $d$ & Pearson $r$ \\\hline
Worldwide & -25.75 & -10.07*** & -1.38 & 0.73*** \\
Argentina & -22.30 & -5.02*** & -0.69 & 0.71*** \\
Australia & -4.96 & -0.86 & -0.12 & 0.65*** \\
Brazil & -0.15 & -0.02 & 0.00 & 0.55*** \\
Canada & -16.03 & -4.30*** & -0.59 & 0.92*** \\
Colombia & -40.62 & -10.21*** & -1.40 & 0.86*** \\
Germany & -19.22 & -5.39*** & -0.74 & 0.64*** \\
India & -10.24 & -2.62* & -0.36 & 0.58*** \\
Indonesia & -50.31 & -6.08*** & -0.84 & 0.79*** \\
Iran & 23.54 & 3.64*** & 0.50 & 0.90*** \\
Italy & -31.03 & -10.97*** & -1.51 & 0.87*** \\
Japan & 1.56 & 0.31 & 0.04 & 0.49*** \\
Mexico & -51.50 & -8.96*** & -1.23 & 0.62*** \\
Nigeria & -54.00 & -8.37*** & -1.15 & 0.36** \\
Pakistan & -42.24 & -10.47*** & -1.44 & 0.38** \\
Philippines & -42.39 & -3.80*** & -0.52 & 0.51*** \\
Russia & -29.98 & -8.77*** & -1.20 & 0.92*** \\
South Korea & -32.86 & -6.14*** & -0.84 & 0.75*** \\
Spain & -12.41 & -3.87*** & -0.53 & 0.91*** \\
Thailand & -36.16 & -6.00*** & -0.82 & 0.76*** \\
Turkey & -37.88 & -3.95*** & -0.54 & 0.66*** \\
Ukraine & -41.42 & -8.04*** & -1.10 & 0.92*** \\
United Kingdom & 5.17 & 1.37 & 0.19 & 0.84*** \\
United States & 6.57 & 2.52* & 0.35 & 0.95*** \\
Vietnam & -22.40 & -1.35 & -0.19 & -0.06 \\
\end{tabular*}
\noindent\rule{\linewidth}{0.6pt}
\end{table}

Table~\ref{tab:language_changes} provides an external validation of the observed trends by examining Wikipedia page-view data across seven major languages. The results indicate a consistent and widespread decline in page-view activity across nearly all analyzed physics-related topics and languages. While the decline observed in English Wikipedia is relatively moderate (averaging approximately 20\%), other languages exhibit substantially sharper decreases. For example, page views associated with topics such as \textit{Kinetic energy} and \textit{Coulomb’s law} declined by around or exceeding 50\% in Hindi, Spanish, Arabic, and Portuguese. A notable exception is the topic \textit{Maxwell’s equations} in English, which shows a 10\% increase in page views, in contrast to the significant declines observed for the same topic in all other examined languages.

\begin{table}[htbp]
\centering
\caption{Percentage change in Wikipedia page views comparing the 2023 reference academic year with the 2025 academic year.  Language abbreviations: \textbf{Eng (English)}, Hin (Hindi), Spa (Spanish), Arb (Arabic), Por (Portuguese), Ita (Italian), and Ger (German). Note the relatively moderate declines in \textbf{English} compared to other languages.}
\label{tab:language_changes}
\setlength{\tabcolsep}{2pt} 
\footnotesize 
\begin{tabular}{lp{0.8cm}p{0.8cm}p{0.8cm}p{0.8cm}p{0.8cm}p{0.8cm}p{0.8cm}}
\hline
Title 
& \centering \textbf{Eng} \%$\Delta$ 
& \centering Hin \%$\Delta$ 
& \centering Spa \%$\Delta$ 
& \centering Arb \%$\Delta$ 
& \centering Por \%$\Delta$ 
& \centering Ita \%$\Delta$ 
& \centering Ger \%$\Delta$ \tabularnewline
\hline
Kinetic energy & \textbf{-24.4} & -72.8 & -56.4 & -57.6 & -60.9 & -35.8 & -17.9 \\
Newton's laws  & \textbf{-22.1} & -59.9 & -55.9 & -58.3 & -58.5 & -11.9 & -8.5 \\
Torque         & \textbf{-42.1} & -33.8 & -55.0 & -57.3 & -55.2 & -42.4 & -31.6 \\
Ang. momentum  & \textbf{-19.1} & -44.4 & -31.5 & -50.7 & -25.1 & -34.3 & -19.7 \\
Maxwell's eq.  & \textbf{10.0}  & -62.0 & -36.5 & -40.7 & -24.4 & -21.5 & -3.6 \\
Electric field &\textbf{-22.7} & -59.5 & -63.7 & -51.6 & -52.4 & -42.6 & -34.6 \\
Electric charge& \textbf{-17.9} & -50.5 & -57.4 & -53.0 & -42.8 & -40.2 & -46.7 \\
Coulomb's law  &\textbf{-20.4} & -66.2 & -46.3 & -53.1 & -55.9 & -53.6 & -22.5 \\
Elec. potential&\textbf{-21.7}& -62.4 & -59.0 & -64.8 & -59.4 & -33.7 & -21.4 \\
\hline
\end{tabular}
\end{table}

\subsection{Cross-Regional Adoption of GenAI in Science}

To evaluate the potential displacement of traditional information-seeking behaviors by generative tools, we conducted a cross-regional analysis of search interest in \textit{ChatGPT} within the \textit{Science} category. The analysis focuses on five representative regions—India, Brazil, Argentina, the United States, and the United Kingdom—and employs the cross-regional normalization defined in Eq.~\ref{eq:cross_regional_rsv}.

\begin{figure}[ht]
\centering
\includegraphics[width=\linewidth]{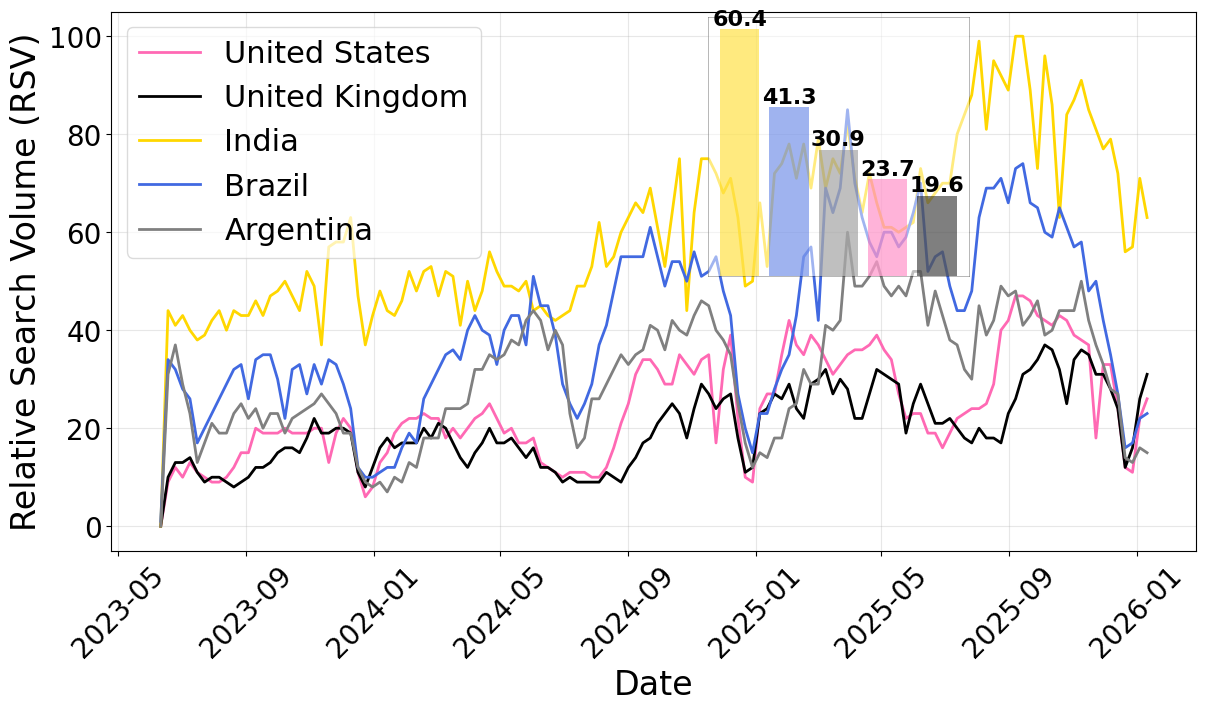}
\caption{Relative Search Volume (RSV) for the topic "ChatGPT" within the "Science" category across five representative regions (2023--2026). The main plot illustrates the longitudinal trends and academic periodicity, while the inset displays the average RSV for each country, highlighting the higher attention density in India and Latin America compared to the US and UK.}
\label{fig:chatgpt_trends}
\end{figure}

The longitudinal data, illustrated in Fig.~\ref{fig:chatgpt_trends}, reveal a profound geographic divergence in the adoption of GenAI for scientific inquiry. High-density regions, most notably India, exhibit the highest relative engagement with a mean RSV of 60.4, followed by Brazil and Argentina with average values of 41.3 and 30.9, respectively. These regions directly correspond to the geographic contexts where the most significant declines in traditional physics search volume were observed. In contrast, English-speaking developed economies, such as the United States and the United Kingdom, demonstrate substantially lower relative interest, with average RSV values of 23.7 and 19.6, respectively. 

Despite these varying intensities, a synchronized "academic pulse" is evident across all five regions. The temporal patterns exhibit sharp contractions in interest during December--January and mid-year periods, coinciding with global academic breaks, followed by rapid recoveries at the onset of new semesters. This periodicity strongly suggests that the utilization of GenAI in these contexts is driven by academic requirements. These findings provide empirical support for the hypothesis that in regions where traditional search activity has sharply declined, there is a commensurate adoption of GenAI as a significant tool for scientific information retrieval.

\section{Discussion \& Conclusion}
This study set out to examine shifts in physics-related information-seeking behavior following the emergence of GenAI. Physics-related search volume and Wikipedia page views have declined significantly across nearly all analyzed topics, while the seasonal academic structure remains intact. Mechanics-related content exhibits larger and more consistent reductions than Electromagnetism. The announced regional heterogeneity is observed: the United States and the United Kingdom show relative stability or moderate declines, while non-English speaking regions, including India, Latin America, and countries where Spanish, Portuguese, and Hindi predominate, exhibit substantially larger reductions, frequently exceeding 50\%. These are also the regions with the highest search volume for GenAI tools in scientific contexts.

\textit{1. How do longitudinal trends in the use of search engines as a physics learning tool change following the emergence of Generative Artificial Intelligence, particularly in terms of search volume and seasonal periodicity?}

The decline in search volume, alongside rising search interest in ChatGPT within the Science category, is consistent with the interpretation that GenAI tools are increasingly being used in ways that overlap with information-seeking practices previously mediated by search engines. This interpretation is consistent with surveys reporting increased GenAI adoption among students globally \cite{microsoft2026global, oecd2025experience}. Notably, the persistence of academic cycling throughout the decline suggests that the underlying learning activity continues; students still engage with physics content according to the semester structure, but the primary mode of information access appears to be shifting from search-based retrieval toward prompt-based generation.

\textit{2. To what extent do changes in search patterns differ between the introductory physics domains of Classical Mechanics and Electromagnetism following the emergence of GenAI?}

Mechanics-related content exhibits larger and more consistent declines in search volume than Electromagnetism across all three regions examined (United States, India, and Worldwide). The distributions of percentage change show that Mechanics-related topics cluster tightly around substantial negative values, whereas Electromagnetism-related topics display greater variability, with some showing stability or modest increases. This pattern is reflected in both the median percentage changes and the standardized effect sizes: Cohen's d values for Mechanics consistently exceed those for Electromagnetism, with the majority surpassing the threshold for a large effect.

This difference may reflect the distinct epistemic structures of the two domains. Mechanics concepts are closely tied to everyday experience and are often approached through verbal causal reasoning \cite{disessa1988knowledge,disessa1993toward,hammer2000student}. By contrast, electromagnetism relies more heavily on abstract constructs such as fields and potentials, which depend strongly on diagrammatic and visual representations \cite{maloney2001surveying,chi1994things,kohl2006effects}. Recent studies have shown that current LLM-based tools are less capable of interpreting and generating such representations \cite{98hg-rkrf,PhysRevPhysEducRes21010154}. Students may therefore be less likely to rely on current GenAI tools for electromagnetism-related learning tasks, where visual reasoning is central, than for mechanics, where verbal explanations are more readily supported.

 \textit{3. How does the magnitude of observed changes in search behavior vary across different countries and cultures?}

There are substantial cross-national differences in the magnitude of search reduction across countries. The United States exhibits a distinct pattern. Mechanics-related search terms show weak and often non-significant declines, while Electromagnetism-related topics remain stable or show slight increases. This contrasts with India and the global aggregate, where both domains exhibit substantial and consistent declines. When examining specific foundational topics across more than 20 countries, the predominantly English-speaking nations, the United States, the United Kingdom, and Australia, consistently exhibit significantly smaller declines than other regions. Wikipedia page view data provide direct support for this pattern: declines in English-language page views are approximately half the magnitude of those observed in other major languages examined. Furthermore, when examining relative search interest in ChatGPT within scientific contexts, attention density in the United States and the United Kingdom is comparable and substantially lower than in India, Brazil, and Argentina.

These regional differences align with recent findings on AI-assisted academic writing. Yu et al. analyzed more than one million submissions of college students' writings and found that the quality gap of the writing between linguistically advantaged and disadvantaged students narrowed after LLM availability \cite{yu2024whose}. Studies of scientific publishing reveal similar patterns: Filimonovic et al. found significant stylistic convergence toward native English writing among 5.65 million articles, with the strongest effects in countries linguistically distant from English \cite{filimonovic2025generative}, and Lin et al. showed that ChatGPT enhanced lexical complexity primarily in abstracts by non-native speakers \cite{lin2025chatgpt}. These studies document increased adoption of AI for content production among non-native English speakers; our findings extend this pattern to content consumption. Consistent with prior observations of the Anglo-centric digital knowledge ecosystem \cite{zuckerman2013rewire}, our results suggest that generative AI may partially mitigate linguistic barriers to accessing information, effectively reducing the 'language tax' and fostering greater equity in knowledge consumption.

These findings also align with broader patterns of AI adoption beyond academic contexts. The Microsoft AI Diffusion Report indicates that the United States ranks 24th globally in AI usage among the working-age population, with trust levels of approximately 32\% compared to 67\% in regions such as the UAE \cite{microsoft2026global}. Lower trust in AI systems may lead users in English-speaking countries to continue to rely on traditional search tools. Of course these are rapidly moving technological and cultural trends. No matter the differences, it appears that GenAI tools and associated queries appear to be an educational practice of students that is here to stay and is likely to grow. 

An informative outlier in the present dataset is Iran, which provides a useful point of comparison for interpreting these regional patterns. Whereas most analyzed countries exhibit substantial declines in physics-related search activity between the reference and integration periods, Iran shows stable or increasing search volumes across multiple foundational topics. This case is particularly informative because Google remains the dominant search platform in Iran, while access to generative AI tools such as ChatGPT is structurally limited \cite{OpenAI2026SupportedCountries}. As a result, Iran approximates a natural counterfactual: a context in which the search infrastructure remains intact while generative alternatives are largely absent. Although the present study does not claim formal causality, this configuration lends additional support to an interpretation linking the global decline in physics-related search activity to the availability of generative tools. Russia represents a different type of outlier. There, Google is not the primary search engine, with a substantial share of search activity occurring on domestic platforms such as Yandex, and Google Trends therefore does not reliably reflect overall search behavior.

\subsection{On the Adoption of GenAI Queries}

While students may be privileging access to information in their native language, which GenAI readily provides, language accessibility alone is unlikely to account for the observed shifts in information-seeking behavior. More fundamentally, GenAI querying differs epistemically from traditional search-based information retrieval. Search typically requires users to formulate effective queries, evaluate the relevance and trustworthiness of multiple sources, and synthesize information across documents, practices that have been shown to require explicit instruction rather than emerge naturally from access alone\cite{caulfield2017}.

In contrast, GenAI systems tend to present responses as coherent and synthesized explanations that appear authoritative and complete. Although some platforms provide references, these are often embedded as follow-up resources rather than serving as the primary objects of evaluation. Previous work has noted that such systems can obscure uncertainty and limitations through fluent language, thus shifting the interpretive burden placed on users \cite{kasneci2023chatgpt}. Moreover, the outcomes of generative queries are highly sensitive to how questions are formulated and iteratively refined, introducing a distinct set of skills associated with effective prompting and critical interpretation.

These differences are not intended as a value judgment, but rather as an indication of a shift in the epistemic skills required for effective information use. As Reich observes, higher education has only recently begun to develop coherent approaches to teaching effective search practices, drawing on work such as Caulfield’s framework for web-based fact checking. Comparable pedagogical frameworks for the use of GenAI remain largely undeveloped \cite{reich2025stop}. The present findings motivate the need for large-scale, locally-grounded, transparent and careful studies of how students engage with generative systems as part of their disciplinary learning practices.

\subsection{Implications for Task Design and Physics Teaching}
Building on the epistemic shift outlined above, we now consider implications for task design and physics teaching. The decline in search-based information seeking, together with the increasing use of GenAI tools in scientific contexts that closely follows academic cycles, indicates that students’ engagement with physics learning persists, even as the means through which explanations are accessed are changing. As explanations are increasingly produced through systems that allow rapid, on-demand generation of responses tailored to students’ immediate questions, physics learning places greater emphasis on a distinct set of skills.

In this learning context, the pedagogical focus shifts from locating explanations to evaluating their quality. Central practices include examining the conceptual coherence of an explanation, checking its consistency with physical principles and representations, identifying omissions or oversimplifications, and recognizing the conditions under which an explanation is partial or limited.

Instructional tasks can make these skills explicit by engaging students directly with generated explanations. For example, students may be asked to analyze an AI-generated explanation of Newton’s third law and identify where it is incomplete or misleading, or to revise the explanation so that it more accurately reflects the relevant physical principles. Such tasks emphasize the development of understanding, judgment and broader meta-cognitive skills, rather than reliance on readily available information.

Comparative tasks offer an additional instructional opportunity. Asking students to compare their own reasoning with a generated response and with a formal solution or a simple representation (such as a diagram or graph) can help distinguish between explanations that sound plausible and those that reflect more robust physical understanding. This distinction is particularly important in topics where fluent verbal descriptions can create a sense of understanding even when conceptual grasp is limited.

Finally, interaction with generative tools can be viewed as a possible extension of an existing syllabus rather than a fundamental curricular shift. This does not require teaching prompt construction as a separate technical skill, but instead highlights familiar practices in physics teaching, such as formulating precise questions, requesting different representations, and examining the scope and limits of an explanation. For instance, students might be asked to reformulate a question in order to obtain an explanation that includes a concrete physical example or a visual representation, thereby framing engagement with generative systems as part of physics teaching rather than as a technological objective in itself.

\subsection{Limitations}

First, the analysis relies on Relative Search Volume rather than absolute counts of search activity. As a result, observed declines in relative search volume do not necessarily indicate absolute decreases in the number of searches, but may also reflect changes in the overall composition of online search activity.

Second, reductions in search activity do not allow for a direct or causal inference that students have switched to GenAI tools. The interpretation linking declining search volume to the adoption of GenAI is based on temporal alignment, parallel patterns observed across datasets, and consistency with external reports, but remains inferential in nature.

Third, the data do not provide insight into how GenAI tools are used, including the depth of interaction, the types of question posed, or the quality of the explanations consumed. Consequently, the study cannot assess whether changes in information-seeking behavior are associated with a deeper or more superficial understanding of the physics.

Finally, the period examined is characterized by rapid technological change in GenAI tools and their modes of use. The observed patterns may therefore reflect a transitional phase in technological adoption and may evolve as these tools, along with educational norms and practices, continue to develop.

\begin{acknowledgments}
The first author thanks Nir Kalush of Google for valuable discussions and insights regarding the Google Trends platform that contributed to this research.
\end{acknowledgments}
\clearpage

\appendix
\section{Detailed Statistical Data}

The tables in this appendix provide the complete statistical breakdown for each physics domain and region analyzed in this study. For all tables in this section, the following metrics and conventions apply:

\begin{itemize}
    \item \textbf{Annual Changes:} $\Delta\%$ reports the relative change in mean weekly Relative Search Volume (RSV) between academic years.
    \item \textbf{Statistical Significance:} Comparisons between the 2023 reference year and the 2025 integration year were conducted using paired-sample $t$-tests. Asterisks denote significance levels: $^*p < 0.05$, $^{**}p < 0.01$, and $^{***}p < 0.001$.
    \item \textbf{Effect Size:} Cohen’s $d$ is calculated for the 2023--2025 difference. Following standard conventions, $|d| > 0.8$ is interpreted as a large effect.
    \item \textbf{Structural Integrity:} Pearson ($r$) and Spearman ($\rho$) correlations assess the stability of the academic seasonal pulse across cycles.
\end{itemize}

\begin{table*}[htbp]
\centering
\caption{\label{tab:worldwide_mechanics_changes}
Annual changes and effect sizes: Mechanics (Worldwide)}

\noindent\rule{\textwidth}{0.6pt}
\begin{tabular*}{\textwidth}{@{\extracolsep{\fill}}lrrrrrrrrr}
\hline
Topic 
& \shortstack{$\Delta\%$\\23$\to$24}
& \shortstack{$\Delta\%$\\24$\to$25}
& \shortstack{$\Delta\%$\\23$\to$25}
& \shortstack{$t$\\23$\to$25}
& \shortstack{$d$\\23$\to$25}
& \shortstack{CI$_{\text{low}}$\\23$\to$25}
& \shortstack{CI$_{\text{high}}$\\23$\to$25}
& \shortstack{Pearson\\$r$\\23$\to$25}
& \shortstack{Spearman\\$\rho$\\23$\to$25} \\
\hline
               Angular frequency & -4.73 & -17.45 & -21.36 & -9.99*** & -1.37 & -12.89 & -8.58 & 0.87*** & 0.86*** \\ 
        Angular momentum & -15.54 & -18.88 & -31.49 & -16.80*** & -2.31 & -21.31 & -16.76 & 0.84*** & 0.83*** \\ 
        Angular velocity & -16.99 & -19.20 & -32.93 & -20.04*** & -2.75 & -23.54 & -19.25 & 0.95*** & 0.94*** \\ 
        Axis of rotation & -53.38 & -0.54 & -53.63 & -10.08*** & -1.39 & -33.75 & -22.55 & 0.73*** & 0.71*** \\ 
        Center of mass & -9.14 & -12.21 & -20.23 & -9.07*** & -1.25 & -14.33 & -9.14 & 0.90*** & 0.86*** \\ 
        Centripetal acceleration & -33.61 & -24.21 & -49.68 & -16.28*** & -2.24 & -34.72 & -27.10 & 0.62*** & 0.64*** \\ 
        Coefficient of friction & -6.27 & -19.13 & -24.20 & -11.37*** & -1.56 & -17.56 & -12.29 & 0.85*** & 0.84*** \\ 
        Elastic collision & -9.10 & -13.64 & -21.50 & -6.27*** & -0.86 & -13.95 & -7.19 & 0.74*** & 0.83*** \\ 
        Force & -16.75 & -14.00 & -28.41 & -20.44*** & -2.81 & -23.54 & -19.33 & 0.92*** & 0.90*** \\ 
        Free body diagram & -3.48 & 14.57 & 10.58 & 3.72*** & 0.51 & 2.41 & 8.04 & 0.88*** & 0.85*** \\ 
        Gravitational energy & -15.73 & -13.61 & -27.20 & -13.05*** & -1.79 & -20.96 & -15.38 & 0.91*** & 0.89*** \\ 
        Inelastic collision & -12.62 & -8.45 & -20.01 & -8.55*** & -1.17 & -15.14 & -9.39 & 0.85*** & 0.79*** \\ 
        Kepler's laws of planetary motion & -21.08 & -14.92 & -32.85 & -12.54*** & -1.72 & -22.96 & -16.63 & 0.87*** & 0.84*** \\ 
        Kinematics & -7.87 & -7.86 & -15.11 & -11.58*** & -1.59 & -11.09 & -7.82 & 0.93*** & 0.91*** \\ 
        Kinetic energy & -15.96 & -14.82 & -28.42 & -13.87*** & -1.91 & -21.88 & -16.35 & 0.92*** & 0.90*** \\ 
        Mechanical energy & -16.95 & -18.66 & -32.45 & -10.90*** & -1.50 & -22.36 & -15.41 & 0.86*** & 0.84*** \\ 
        Moment of inertia & -12.93 & -12.19 & -23.55 & -15.50*** & -2.13 & -14.28 & -11.00 & 0.91*** & 0.87*** \\ 
        Momentum & -13.51 & -16.80 & -28.04 & -16.84*** & -2.31 & -23.84 & -18.76 & 0.85*** & 0.80*** \\ 
        Newton's law of universal gravitation & -22.82 & -19.05 & -37.52 & -15.41*** & -2.12 & -25.74 & -19.81 & 0.94*** & 0.95*** \\ 
        Newton's laws of motion & -20.89 & -23.94 & -39.83 & -17.29*** & -2.37 & -30.39 & -24.07 & 0.85*** & 0.85*** \\ 
        Newton's second law & -8.30 & -15.87 & -22.85 & -14.12*** & -1.94 & -16.14 & -12.12 & 0.91*** & 0.90*** \\ 
        Normal Force & -13.47 & -11.77 & -23.65 & -10.17*** & -1.40 & -15.59 & -10.45 & 0.87*** & 0.92*** \\ 
        Parallel axis theorem & -17.80 & -9.40 & -25.52 & -7.19*** & -0.99 & -14.43 & -8.14 & 0.72*** & 0.82*** \\ 
        Potential energy & -10.34 & -14.68 & -23.50 & -11.29*** & -1.55 & -17.38 & -12.13 & 0.87*** & 0.84*** \\ 
        Projectile motion & -19.13 & -22.50 & -37.32 & -16.02*** & -2.20 & -27.88 & -21.67 & 0.81*** & 0.77*** \\ 
        Relative velocity & -18.43 & -12.23 & -28.41 & -16.06*** & -2.21 & -17.11 & -13.31 & 0.89*** & 0.77*** \\ 
        Restoring force & -11.34 & -22.08 & -30.91 & -9.77*** & -1.34 & -20.40 & -13.45 & 0.87*** & 0.88*** \\ 
        Simple harmonic motion & -17.99 & -16.14 & -31.22 & -9.23*** & -1.27 & -19.02 & -12.23 & 0.83*** & 0.84*** \\ 
        Tangential acceleration & -6.18 & -10.93 & -16.44 & -5.60*** & -0.77 & -11.18 & -5.28 & 0.84*** & 0.85*** \\ 
        Torque & -20.86 & -10.82 & -29.42 & -15.53*** & -2.13 & -22.69 & -17.50 & 0.77*** & 0.79***  \\ \hline
\end{tabular*}

\noindent\rule{\textwidth}{0.6pt}
\end{table*}

\begin{table*}[htbp]
\centering
\caption{\label{tab:worldwide_em_changes}
Annual changes and effect sizes: Electromagnetism (Worldwide).}

\noindent\rule{\textwidth}{0.6pt}

\begin{tabular*}{\textwidth}{@{\extracolsep{\fill}}lrrrrrrrrr}
\hline
Topic 
& \shortstack{$\Delta\%$\\23$\to$24}
& \shortstack{$\Delta\%$\\24$\to$25}
& \shortstack{$\Delta\%$\\23$\to$25}
& \shortstack{$t$\\23$\to$25}
& \shortstack{$d$\\23$\to$25}
& \shortstack{CI$_{\text{low}}$\\23$\to$25}
& \shortstack{CI$_{\text{high}}$\\23$\to$25}
& \shortstack{Pearson\\$r$}
& \shortstack{Spearman\\$\rho$} \\
\hline
\\ 
        Ampère's circuital law & -13.63 & -6.48 & -19.22 & -6.28*** & -0.86 & -12.57 & -6.48 & 0.58*** & 0.71*** \\ 
        Biot–Savart law & -19.84 & -11.41 & -28.99 & -8.20*** & -1.13 & -17.57 & -10.66 & 0.38* & 0.49*** \\ 
        Capacitor & -11.07 & -16.70 & -25.92 & -13.30*** & -1.83 & -19.78 & -14.60 & 0.74*** & 0.85*** \\ 
        Coulomb's law & -21.78 & -16.52 & -34.70 & -13.79*** & -1.89 & -24.73 & -18.44 & 0.79*** & 0.69*** \\ 
        Current loop & -10.83 & 41.21 & 25.92 & 3.11** & 0.43 & 1.79 & 8.32 & 0.45*** & 0.51*** \\ 
        Displacement current & -10.48 & -12.10 & -21.31 & -3.52*** & -0.48 & -9.15 & -2.51 & 0.57*** & 0.80*** \\ 
        Electric charge & -15.45 & -12.09 & -25.67 & -15.47*** & -2.12 & -20.34 & -15.66 & 0.92*** & 0.88*** \\ 
        Electric current & -16.37 & -16.00 & -29.75 & -9.79*** & -1.34 & -18.30 & -12.07 & 0.93*** & 0.91*** \\ 
        Electric dipole & 135.22 & 36.15 & 220.26 & 12.11*** & 1.66 & 19.00 & 26.55 & 0.31* & 0.21 \\ 
        Electric field & -12.87 & -14.77 & -25.74 & -10.07*** & -1.38 & -20.00 & -13.36 & 0.73*** & 0.89*** \\ 
        Electric flux & -19.79 & -18.86 & -34.92 & -11.84*** & -1.63 & -21.49 & -15.26 & 0.66*** & 0.85*** \\ 
        Electric potential & -15.32 & -17.72 & -30.33 & -16.89*** & -2.32 & -22.27 & -17.54 & 0.75*** & 0.90*** \\ 
        Electric potential energy & -10.37 & -13.86 & -22.79 & -8.29*** & -1.14 & -16.64 & -10.15 & 0.73*** & 0.80*** \\ 
        Electrical impedance & -1.15 & -3.14 & -4.25 & -1.86 & -0.26 & -6.39 & 0.24 & 0.29* & 0.35* \\ 
        Equipotential & -14.57 & -6.33 & -19.98 & -4.90*** & -0.67 & -10.90 & -4.57 & 0.69*** & 0.84*** \\ 
        Equipotential surface & -18.98 & 13.43 & -8.09 & -1.24 & -0.17 & -10.45 & 2.45 & 0.34* & 0.39** \\ 
        Faraday's law of induction & -17.09 & -12.81 & -27.70 & -12.42*** & -1.71 & -18.30 & -13.21 & 0.79*** & 0.85*** \\ 
        Gauss's law & -14.81 & -13.74 & -26.52 & -8.16*** & -1.12 & -16.46 & -9.96 & 0.58*** & 0.81*** \\ 
        Inductance & -28.11 & -19.45 & -42.09 & -7.23*** & -0.99 & -21.55 & -12.19 & 0.53*** & 0.74*** \\ 
        Inductor & -9.61 & -15.07 & -23.23 & -9.04*** & -1.24 & -16.67 & -10.61 & 0.78*** & 0.78*** \\ 
        LC circuit & -45.43 & -61.44 & -78.96 & -10.48*** & -1.44 & -30.28 & -20.55 & 0.42* & 0.53*** \\ 
        Lenz's law & -14.25 & -10.55 & -23.29 & -7.87*** & -1.08 & -13.78 & -8.18 & 0.72*** & 0.78*** \\ 
        Lorentz force & -28.37 & -17.14 & -40.64 & -7.76*** & -1.07 & -22.49 & -13.25 & 0.73*** & 0.83*** \\ 
        Magnetic energy & -8.99 & -2.95 & -11.68 & -3.08** & -0.42 & -8.67 & -1.82 & 0.46*** & 0.55*** \\ 
        Magnetic field & -12.89 & -7.25 & -19.21 & -8.26*** & -1.13 & -12.62 & -7.68 & 0.80*** & 0.89*** \\ 
        Magnetic force & -14.06 & -17.81 & -29.37 & -7.69*** & -1.06 & -19.49 & -11.42 & 0.81*** & 0.81*** \\ 
        Magnetic levitation & -1.62 & 17.14 & 15.24 & 2.89** & 0.40 & 2.11 & 11.74 & 0.28* & 0.28* \\ 
        Maxwell’s equations & -12.80 & -8.96 & -20.62 & -14.30*** & -1.96 & -18.50 & -13.95 & 0.78*** & 0.78*** \\ 
        Ohm's law & -17.22 & -19.38 & -33.27 & -15.91*** & -2.19 & -25.77 & -20.00 & 0.77*** & 0.82*** \\ 
        RC circuit & -21.29 & -29.76 & -44.71 & -10.79*** & -1.48 & -31.86 & -21.87 & 0.45*** & 0.45*** \\ 
        Resistor & -33.44 & -7.24 & -38.26 & -8.62*** & -1.18 & -22.52 & -14.01 & 0.76*** & 0.82*** \\ 
        RL circuit & -16.46 & -28.96 & -40.65 & -10.09*** & -1.39 & -28.43 & -19.00 & 0.58*** & 0.56*** \\
\hline
\end{tabular*}

\noindent\rule{\textwidth}{0.6pt}
\end{table*}

\begin{table*}[htbp]
\centering
\caption{\label{tab:us_mechanics_changes}
Annual changes and effect sizes: Mechanics (United States).}

\noindent\rule{\textwidth}{0.6pt}

\begin{tabular*}{\textwidth}{@{\extracolsep{\fill}}lrrrrrrrrr}
\hline
Topic 
& \shortstack{$\Delta\%$\\23$\to$24}
& \shortstack{$\Delta\%$\\24$\to$25}
& \shortstack{$\Delta\%$\\23$\to$25}
& \shortstack{$t$\\23$\to$25}
& \shortstack{$d$\\23$\to$25}
& \shortstack{CI$_{\text{low}}$\\23$\to$25}
& \shortstack{CI$_{\text{high}}$\\23$\to$25}
& \shortstack{Pearson\\$r$\\23$\to$25}
& \shortstack{Spearman\\$\rho$\\23$\to$25} \\
\hline
        Angular frequency & -6.59 & -7.48 & -13.58 & -2.71** & -0.37 & -8.80 & -1.32 & 0.85*** & 0.87*** \\ 
        Angular momentum & 2.41 & -5.94 & -3.67 & -0.86 & -0.12 & -4.96 & 1.98 & 0.87*** & 0.88*** \\ 
        Angular velocity & -0.40 & -9.36 & -9.71 & -2.79** & -0.38 & -7.95 & -1.30 & 0.90*** & 0.88*** \\ 
        Center of mass & 6.22 & -4.45 & 1.49 & 0.48 & 0.07 & -2.16 & 3.51 & 0.90*** & 0.88*** \\ 
        Centripetal acceleration & -40.53 & -25.98 & -55.98 & -9.04*** & -1.24 & -30.41 & -19.36 & 0.63*** & 0.67*** \\ 
        Coefficient of friction & 4.49 & -3.22 & 1.12 & 0.32 & 0.04 & -2.37 & 3.27 & 0.87*** & 0.88*** \\ 
        Elastic collision & -10.16 & -9.87 & -19.02 & -3.66*** & -0.50 & -13.79 & -4.02 & 0.85*** & 0.81*** \\ 
        Force & -4.04 & -6.67 & -10.44 & -3.55*** & -0.49 & -10.16 & -2.82 & 0.87*** & 0.86*** \\ 
        Free body diagram & 3.91 & 17.89 & 22.50 & 4.67*** & 0.64 & 4.09 & 10.25 & 0.92*** & 0.92*** \\ 
        Gravitational energy & -13.45 & -3.46 & -16.45 & -3.71*** & -0.51 & -13.08 & -3.90 & 0.83*** & 0.80*** \\ 
        Inelastic collision & -6.01 & -2.57 & -8.43 & -1.49 & -0.21 & -8.80 & 1.29 & 0.83*** & 0.79*** \\ 
        Kepler's laws of planetary motion & -6.43 & -5.88 & -11.92 & -2.77** & -0.38 & -10.02 & -1.60 & 0.83*** & 0.82*** \\ 
        Kinematics & -0.61 & -0.31 & -0.92 & -0.38 & -0.05 & -2.14 & 1.46 & 0.95*** & 0.93*** \\ 
        Kinetic energy & -7.16 & 4.35 & -3.12 & -0.98 & -0.13 & -5.47 & 1.88 & 0.87*** & 0.83*** \\ 
        Mechanical energy & -1.49 & -5.82 & -7.23 & -1.32 & -0.18 & -8.77 & 1.82 & 0.78*** & 0.71*** \\ 
        Moment of inertia & 0.68 & -0.27 & 0.41 & 0.11 & 0.02 & -2.92 & 3.26 & 0.92*** & 0.94*** \\ 
        Momentum & -4.87 & -7.76 & -12.25 & -3.91*** & -0.54 & -11.57 & -3.72 & 0.84*** & 0.83*** \\ 
        Newton's law of universal gravitation & -1.33 & 0.82 & -0.53 & -0.12 & -0.02 & -5.14 & 4.57 & 0.76*** & 0.70*** \\ 
        Newton's laws of motion & -23.43 & -16.00 & -35.68 & -7.09*** & -0.97 & -22.41 & -12.53 & 0.80*** & 0.78*** \\ 
        Newton's second law & 0.64 & -6.79 & -6.19 & -1.57 & -0.22 & -6.63 & 0.82 & 0.88*** & 0.88*** \\ 
        Normal Force & 2.93 & -1.23 & 1.67 & 0.36 & 0.05 & -2.52 & 3.61 & 0.92*** & 0.93*** \\ 
        Parallel axis theorem & -8.68 & 22.36 & 11.74 & 1.52 & 0.21 & -1.12 & 8.07 & 0.80*** & 0.84*** \\ 
        Potential energy & 7.05 & -6.03 & 0.59 & 0.18 & 0.02 & -3.55 & 4.23 & 0.88*** & 0.82*** \\ 
        Projectile motion & -4.43 & -9.31 & -13.32 & -4.17*** & -0.57 & -7.74 & -2.71 & 0.94*** & 0.90*** \\ 
        Relative velocity & -5.82 & -8.27 & -13.61 & -2.98** & -0.41 & -7.61 & -1.49 & 0.86*** & 0.84*** \\ 
        Simple harmonic motion & -3.94 & -2.19 & -6.04 & -1.51 & -0.21 & -4.78 & 0.67 & 0.91*** & 0.92*** \\ 
        Torque & -0.72 & 5.46 & 4.69 & 1.62 & 0.22 & -0.61 & 5.74 & 0.87*** & 0.86*** \\ 
\hline
\end{tabular*}

\noindent\rule{\textwidth}{0.6pt}
\end{table*}

\begin{table*}[htbp]
\centering
\caption{\label{tab:us_em_changes}
Annual changes and effect sizes: Electromagnetism (United States).}

\noindent\rule{\textwidth}{0.6pt}

\begin{tabular*}{\textwidth}{@{\extracolsep{\fill}}lrrrrrrrrr}
\hline
Topic 
& \shortstack{$\Delta\%$\\23$\to$24}
& \shortstack{$\Delta\%$\\24$\to$25}
& \shortstack{$\Delta\%$\\23$\to$25}
& \shortstack{$t$\\23$\to$25}
& \shortstack{$d$\\23$\to$25}
& \shortstack{CI$_{\text{low}}$\\23$\to$25}
& \shortstack{CI$_{\text{high}}$\\23$\to$25}
& \shortstack{Pearson\\$r$}
& \shortstack{Spearman\\$\rho$} \\
\hline
        Ampère's circuital law & 15.26 & 10.69 & 27.59 & 3.88*** & 0.53 & 3.42 & 10.77 & 0.84*** & 0.84*** \\ 
        Capacitor & 6.26 & -7.51 & -1.72 & -0.57 & -0.08 & -3.15 & 1.76 & 0.90*** & 0.88*** \\ 
        Coulomb's law & 0.51 & 6.64 & 7.19 & 2.28* & 0.31 & 0.38 & 5.96 & 0.90*** & 0.89*** \\ 
        Electric charge & -0.98 & -10.04 & -10.92 & -3.77*** & -0.52 & -8.36 & -2.55 & 0.92*** & 0.92*** \\ 
        Electric field & 8.43 & -1.71 & 6.57 & 2.52* & 0.35 & 0.45 & 3.97 & 0.95*** & 0.93*** \\ 
        Electric potential & 7.29 & -1.62 & 5.55 & 1.96 & 0.27 & -0.03 & 3.28 & 0.96*** & 0.93*** \\ 
        Electric potential energy & 5.55 & -5.26 & 0.00 & 0.00 & 0.00 & -2.79 & 2.79 & 0.91*** & 0.87*** \\ 
        Electrical impedance & 21.27 & 4.80 & 27.09 & 4.56*** & 0.63 & 5.95 & 15.30 & 0.40** & 0.25 \\ 
        Equipotential & 5.84 & 4.54 & 10.65 & 1.44 & 0.20 & -0.83 & 5.09 & 0.87*** & 0.81*** \\ 
        Faraday's law of induction & 8.24 & 5.36 & 14.04 & 3.11** & 0.43 & 1.62 & 7.51 & 0.88*** & 0.88*** \\ 
        Gauss's law & 5.48 & 1.78 & 7.36 & 1.69 & 0.23 & -0.37 & 4.37 & 0.91*** & 0.87*** \\ 
        Lenz's law & 15.40 & -2.17 & 12.90 & 1.80 & 0.25 & -0.35 & 6.39 & 0.80*** & 0.76*** \\ 
        Lorentz force & -5.70 & 3.02 & -2.85 & -0.48 & -0.07 & -6.89 & 4.21 & 0.70*** & 0.68*** \\ 
        Magnetic field & 6.62 & -0.70 & 5.87 & 1.94 & 0.27 & -0.10 & 6.29 & 0.88*** & 0.87*** \\ 
        Maxwell’s equations & 2.56 & -0.96 & 1.58 & 0.50 & 0.07 & -2.38 & 3.96 & 0.55*** & 0.54*** \\ 
        Ohm's law & 7.27 & -4.77 & 2.15 & 0.67 & 0.09 & -2.16 & 4.31 & 0.82*** & 0.81*** \\ 
        Resistor & -35.18 & 8.66 & -29.56 & -5.22*** & -0.72 & -18.16 & -8.07 & 0.70*** & 0.72*** \\ 
\hline
\end{tabular*}

\noindent\rule{\textwidth}{0.6pt}
\end{table*}

\begin{table*}[htbp]
\centering
\caption{\label{tab:india_mechanics_changes}
Annual changes and effect sizes: Mechanics (India).}

\noindent\rule{\textwidth}{0.6pt}

\begin{tabular*}{\textwidth}{@{\extracolsep{\fill}}lrrrrrrrrr}
\hline
Topic 
& \shortstack{$\Delta\%$\\23$\to$24}
& \shortstack{$\Delta\%$\\24$\to$25}
& \shortstack{$\Delta\%$\\23$\to$25}
& \shortstack{$t$\\23$\to$25}
& \shortstack{$d$\\23$\to$25}
& \shortstack{CI$_{\text{low}}$\\23$\to$25}
& \shortstack{CI$_{\text{high}}$\\23$\to$25}
& \shortstack{Pearson\\$r$\\23$\to$25}
& \shortstack{Spearman\\$\rho$\\23$\to$25} \\
\hline
        Angular frequency & -8.18 & -11.63 & -18.86 & -6.86*** & -0.94 & -13.44 & -7.36 & 0.80*** & 0.77*** \\ 
        Angular momentum & -16.98 & -9.23 & -24.65 & -12.49*** & -1.72 & -17.54 & -12.69 & 0.72*** & 0.71*** \\ 
        Angular velocity & -22.52 & -7.97 & -28.70 & -12.15*** & -1.67 & -19.41 & -13.91 & 0.67*** & 0.73*** \\ 
        Center of mass & -7.71 & -4.00 & -11.40 & -2.13* & -0.29 & -11.10 & -0.33 & 0.65*** & 0.74*** \\ 
        Centripetal acceleration & -60.98 & -34.07 & -74.28 & -12.18*** & -1.67 & -37.82 & -27.12 & 0.33* & 0.28* \\ 
        Coefficient of friction & -17.75 & -17.62 & -32.24 & -10.25*** & -1.41 & -26.35 & -17.72 & 0.25 & 0.23 \\ 
        Elastic collision & -22.39 & -20.37 & -38.20 & -5.96*** & -0.82 & -18.28 & -9.08 & 0.65*** & 0.64*** \\ 
        Force & -23.41 & -7.82 & -29.39 & -17.42*** & -2.39 & -24.28 & -19.27 & 0.63*** & 0.66*** \\ 
        Free body diagram & -18.21 & -11.13 & -27.32 & -5.45*** & -0.75 & -20.44 & -9.44 & 0.21 & 0.27 \\ 
        Gravitational energy & -31.51 & -19.57 & -44.91 & -7.38*** & -1.01 & -27.02 & -15.47 & 0.68*** & 0.69*** \\ 
        Kepler's laws of planetary motion & -27.09 & -19.44 & -41.26 & -12.27*** & -1.69 & -24.41 & -17.55 & 0.85*** & 0.86*** \\
        Kinematics & -9.80 & -10.62 & -19.38 & -10.97*** & -1.51 & -16.16 & -11.16 & 0.76*** & 0.72*** \\ 
        Kinetic energy & -13.33 & -5.10 & -17.75 & -8.23*** & -1.13 & -12.91 & -7.85 & 0.72*** & 0.82*** \\ 
        Mechanical energy & -37.38 & 1.03 & -36.73 & -7.17*** & -0.98 & -19.23 & -10.81 & 0.77*** & 0.82*** \\ 
        Moment of inertia & -13.52 & -7.48 & -19.99 & -5.68*** & -0.78 & -13.89 & -6.64 & 0.73*** & 0.77*** \\ 
        Momentum & -19.01 & -11.49 & -28.32 & -19.11*** & -2.62 & -21.89 & -17.73 & 0.81*** & 0.80*** \\ 
        Newton's law of universal gravitation & -25.61 & -11.51 & -34.17 & -10.37*** & -1.42 & -18.60 & -12.57 & 0.91*** & 0.89*** \\ 
        Newton's laws of motion & -26.47 & -12.98 & -36.01 & -17.73*** & -2.44 & -23.54 & -18.76 & 0.89*** & 0.89*** \\ 
        Newton's second law & -20.37 & -19.20 & -35.66 & -19.24*** & -2.64 & -19.73 & -16.00 & 0.91*** & 0.91*** \\ 
        Normal Force & -42.89 & -41.93 & -66.83 & -4.50*** & -0.62 & -25.51 & -9.77 & 0.20 & 0.21 \\ 
        Parallel axis theorem & -32.82 & -16.67 & -44.02 & -5.43*** & -0.75 & -13.41 & -6.17 & 0.60*** & 0.72*** \\ 
        Potential energy & -11.73 & -5.62 & -16.69 & -4.80*** & -0.66 & -10.09 & -4.14 & 0.50*** & 0.77*** \\ 
        Projectile motion & -21.99 & -14.43 & -33.25 & -12.82*** & -1.76 & -22.69 & -16.55 & 0.84*** & 0.85*** \\ 
        Relative velocity & -31.11 & -12.25 & -39.55 & -14.16*** & -1.94 & -25.44 & -19.12 & 0.80*** & 0.78*** \\ 
        Simple harmonic motion & -7.82 & -1.09 & -8.83 & -2.14* & -0.29 & -5.78 & -0.18 & 0.89*** & 0.92*** \\ 
        Torque & -18.56 & -8.88 & -25.79 & -8.19*** & -1.12 & -14.33 & -8.69 & 0.49*** & 0.61*** \\ 
\hline
\end{tabular*}

\noindent\rule{\textwidth}{0.6pt}
\end{table*}

\begin{table*}[htbp]
\centering
\caption{\label{tab:india_em_changes}
Annual changes and effect sizes: Electromagnetism (India).}

\noindent\rule{\textwidth}{0.6pt}

\begin{tabular*}{\textwidth}{@{\extracolsep{\fill}}lrrrrrrrrr}
\hline
Topic 
& \shortstack{$\Delta\%$\\23$\to$24}
& \shortstack{$\Delta\%$\\24$\to$25}
& \shortstack{$\Delta\%$\\23$\to$25}
& \shortstack{$t$\\23$\to$25}
& \shortstack{$d$\\23$\to$25}
& \shortstack{CI$_{\text{low}}$\\23$\to$25}
& \shortstack{CI$_{\text{high}}$\\23$\to$25}
& \shortstack{Pearson\\$r$\\23$\to$25}
& \shortstack{Spearman\\$\rho$\\23$\to$25} \\
\hline
        Ampère's circuital law & -20.76 & -2.08 & -22.41 & -3.52*** & -0.48 & -10.07 & -2.76 & 0.53*** & 0.78*** \\ 
        Biot–Savart law & -25.43 & -9.93 & -32.84 & -5.23*** & -0.72 & -14.36 & -6.40 & 0.49*** & 0.74*** \\ 
        Capacitor & -18.27 & -5.68 & -22.91 & -5.04*** & -0.69 & -12.50 & -5.38 & 0.39** & 0.77*** \\ 
        Coulomb's law & -26.35 & -8.41 & -32.54 & -9.23*** & -1.27 & -18.83 & -12.11 & 0.55*** & 0.73*** \\ 
        Current loop & -13.75 & 86.65 & 61.00 & 3.56*** & 0.49 & 2.92 & 10.47 & 0.32* & 0.54*** \\ 
        Displacement current & -8.33 & 1.77 & -6.71 & -0.74 & -0.10 & -4.60 & 2.11 & 0.56*** & 0.82*** \\ 
        Electric charge & -18.76 & -2.70 & -20.96 & -6.91*** & -0.95 & -15.09 & -8.30 & 0.46*** & 0.71*** \\ 
        Electric field & -9.46 & -0.87 & -10.24 & -2.62* & -0.36 & -7.82 & -1.04 & 0.58*** & 0.76*** \\ 
        Electric flux & -20.42 & -10.10 & -28.46 & -6.18*** & -0.85 & -16.20 & -8.25 & 0.47*** & 0.72*** \\ 
        Electric potential & -13.22 & -3.34 & -16.12 & -4.15*** & -0.57 & -12.15 & -4.23 & 0.40** & 0.62*** \\ 
        Electric potential energy & -17.41 & -4.93 & -21.48 & -3.82*** & -0.53 & -14.87 & -4.64 & 0.34* & 0.56*** \\ 
        Electrical impedance & -3.80 & -14.19 & -17.44 & -3.61*** & -0.50 & -12.95 & -3.69 & 0.46*** & 0.56*** \\ 
        Equipotential & -22.44 & -3.34 & -25.03 & -3.83*** & -0.53 & -11.68 & -3.64 & 0.38** & 0.70*** \\ 
        Faraday's law of induction & -21.38 & -12.52 & -31.23 & -7.39*** & -1.01 & -15.52 & -8.89 & 0.70*** & 0.81*** \\ 
        Gauss's law & -16.82 & -11.10 & -26.05 & -4.42*** & -0.61 & -12.54 & -4.71 & 0.36** & 0.61*** \\ 
        Inductance & -12.25 & -11.00 & -21.90 & -3.79*** & -0.52 & -9.49 & -2.92 & 0.60*** & 0.83*** \\ 
        Inductor & -20.06 & -13.02 & -30.47 & -4.94*** & -0.68 & -17.00 & -7.18 & 0.59*** & 0.67*** \\ 
        Lenz's law & -18.15 & -12.39 & -28.29 & -4.51*** & -0.62 & -12.24 & -4.70 & 0.61*** & 0.80*** \\ 
        Lorentz force & -27.34 & -15.26 & -38.43 & -5.94*** & -0.82 & -15.22 & -7.53 & 0.41** & 0.64*** \\ 
        Magnetic field & -8.89 & 3.68 & -5.53 & -0.94 & -0.13 & -5.25 & 1.89 & 0.48*** & 0.90*** \\ 
        Maxwell’s equations & -24.85 & -15.05 & -36.16 & -13.29*** & -1.83 & -25.00 & -18.44 & 0.64*** & 0.63*** \\ 
        Ohm's law & -21.19 & -14.94 & -32.96 & -6.45*** & -0.89 & -16.20 & -8.51 & 0.47*** & 0.79*** \\ 
        Resistor & -51.64 & 5.21 & -49.12 & -5.77*** & -0.79 & -16.35 & -7.92 & 0.52*** & 0.86*** \\
        \hline
\end{tabular*}

\noindent\rule{\textwidth}{0.6pt}
\end{table*}

\clearpage
\bibliography{ref}

\end{document}